\documentclass[twocolumn,aps,pra,superscriptaddress,10pt]{revtex4-2}
\usepackage{booktabs}
\usepackage{enumitem}
\usepackage{graphicx}
\usepackage{amsmath}
\usepackage{amssymb}
\usepackage{siunitx}
\usepackage{dcolumn}
\usepackage{bm}
\usepackage{hyperref}
\hypersetup{colorlinks=true,linkcolor=green,citecolor=red,urlcolor=red}
\usepackage{float}
\usepackage{tikz}
\usepackage{multirow}
\usepackage{placeins}

\usepackage[acronym]{glossaries}
\glsdisablehyper
\makeglossaries
\newacronym{qd}{QD}{quantum dot}
\newacronym{cmos}{CMOS}{complementary metal–oxide–semiconductor}
\newacronym{ml}{ML}{machine learning}
\newacronym{cnn}{CNN}{convolutional neural network}
\newacronym{nn}{NN}{neural network}
\newacronym{csd}{CSD}{charge stability diagram}
\newacronym{set}{SET}{single electron transistor}
\newacronym{snr}{SNR}{signal to noise ratio}
\newacronym{fd-soi}{FD-SOI}{fully depleted silicon on insulator}
\newacronym{IoU}{IoU}{intersection over union}
\newacronym{dl}{DL}{deep learning}
\newacronym{fft}{FFT}{Fast Fourier Transform}

\begin{document}
\pagestyle{plain}

\title{Automatic Charge State Tuning of 300 mm FDSOI Quantum Dots Using Neural Network Segmentation of Charge Stability Diagram}

\author{Peter Samaha}
\email{petersamaha25@gmail.com}
\author{Amine Torki}
\author{Ysaline Renaud}
\author{Sam Fiette}
\author{Emmanuel Chanrion}
\author{Pierre-André Mortemousque}
\email{pierre-andre.mortemousque@cea.fr}
\author{Yann Beilliard}
\email{yann.beilliard@cea.fr }
\affiliation{CEA-Leti, Univ. Grenoble Alpes, F-38000 Grenoble, France}
\date{\today}

\begin{abstract}
Tuning of gate-defined semiconductor quantum dots (QDs) is a major bottleneck for scaling spin-qubit technologies. We present a deep learning (DL) driven, semantic-segmentation pipeline that performs charge auto-tuning by locating transition lines in full charge stability diagrams (CSDs) and returns gate voltage targets for the single charge regime. We assemble and manually annotate a large, heterogeneous dataset of 1015 experimental CSDs measured from silicon QD devices, spanning nine design geometries, multiple wafers, and fabrication runs. A U-Net style \gls{cnn} with a MobileNetV2 encoder is trained and validated through five-fold group cross-validation. Our model achieves an overall offline tuning success of 80.0\% in locating the single-charge regime, with peak performance exceeding 88\% for some designs. We analyze dominant failure modes and propose targeted mitigations. Finally, wide-range diagram segmentation also naturally enables scalable physic-based feature extraction that can feed back to fabrication and design workflows and outline a roadmap for real-time integration in a cryogenic wafer prober. Overall, our results show that \gls{nn} based wide-diagram segmentation is a practical step toward automated, high-throughput charge tuning for silicon QD qubits.
\end{abstract}

\maketitle

\section{Introduction}

Quantum computing offers a promising advantage for specific applications, particularly in quantum simulation, cryptography, and optimization \cite{Feynman1982,Shor1994,Grover1997}. This promise rests on our ability to precisely control and manipulate individual quantum systems, a capability that defines the second quantum revolution \cite{macfarlane_quantum_2003}. Translating laboratory demonstrations of such control into practical, large-scale quantum processors is therefore primarily an engineering challenge. Consequently, overcoming these practical obstacles is a necessary step toward building usable and scalable quantum systems. In this paper, we address one such bottleneck, the manual and time-consuming charge tuning of gate-defined semiconductors \glspl{qd}.

\begin{figure*}[h!bt]
    \centering
    \includegraphics[width=1\linewidth]{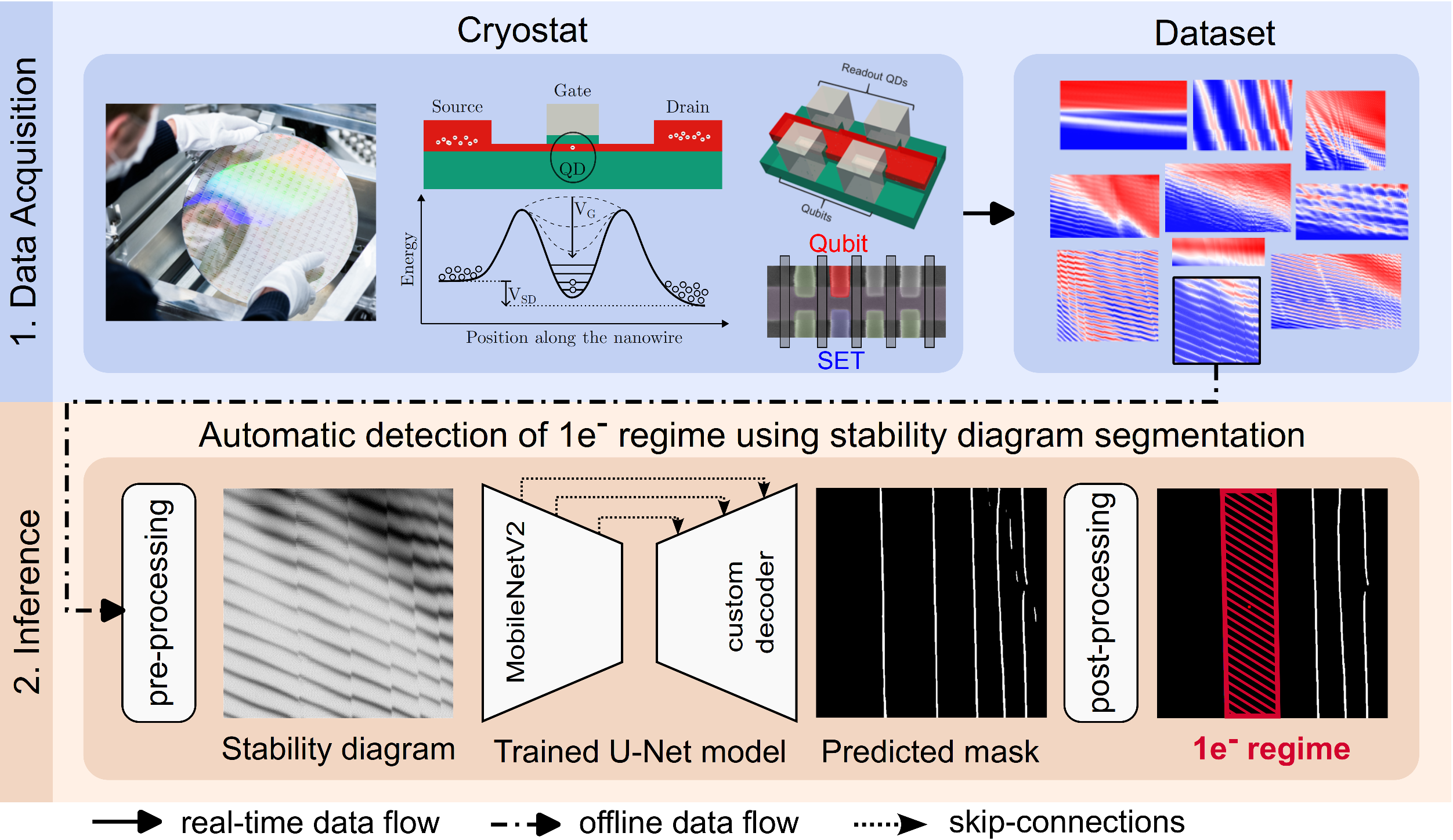}
    \caption{Schematic summary of the offline auto-tuning pipeline. \textbf{Top (Data acquisition):} experimental setup and device illustrations (left) show the measurements done using cryogenic wafer prober on the gate-defined FDSOI \gls{qd} geometry; the dataset panel (right) displays representative measured \glspl{csd} used for training. \textbf{Bottom (Inference):} the wide-diagram segmentation flow. Preprocessing produces a normalized input, which is fed to a U-Net with a MobileNetV2 encoder; the network outputs a predicted probability map that is binarized and refined by morphological post-processing; the algorithm then extracts transition-line skeletons and identifies the single-electron (1e$^-$) regime (highlighted in red here), returning gate-voltage targets for automatic charge tuning. The pipeline is trained and validated on a large, heterogeneous set of experimental \glspl{csd} and produces per-pixel line maps that enable both tuning and physics-based feature extraction.}
    \label{fig:Summary}
\end{figure*}

In this work, we focus on spin qubit implemented in a gate-defined semiconductor \glspl{qd} \cite{Los98,veldhorst_two-qubit_2015,burkard_semiconductor_2023}. More specifically, silicon nanowire \gls{cmos} transistors fabricated on \gls{fd-soi} wafers \cite{maurand_cmos_2016} (device schematics shown in Fig.~\ref{fig:Summary}, top-left). In these devices, the spin state of a single confined electron or hole encodes a quantum bit (qubit). This platform benefits from a range of advantages. For instance, it is compatible with 300 mm \gls{cmos} fabrication methods \cite{maurand_cmos_2016,gonzalez2021,elsayed_low_2024,steinacker2024300}, and such qubit implementation offers long coherence time \cite{veldhorst_addressable_2014, tyryshkin_electron_2012,muhonen_storing_2014}, high gate fidelity \cite{xue_quantum_2022,yoneda_quantum-dot_2018,mills_two-qubit_2022}, and can be operated at moderate cryogenic temperatures of around $1K$ \cite{Huang_2024,petit_design_2022}.

Despite these advantages, the operation of \gls{qd} spin qubit devices faces two big hurdles hindering the development of large-scale quantum computers. First, tuning these devices to the desired charge and tunnel regimes is cumbersome and remains a largely manual process in which experts rely on informed guesses and visual heuristics to adjust the multiple gate voltages to reach the desired configuration. This process is time-consuming, prone to errors, and becomes increasingly complex as the number of gate electrodes grows with array size, making the parameter space too large and impossible to tune manually.
Second, the fabrication of high-quality, reproducible semiconductors \glspl{qd}, where variabilities are well understood, remains challenging. As a result, inevitable device-to-device variability persists across mask layouts, wafers, and process runs \cite{cifuentes2024bounds}, which in turn exacerbates the first challenge by invalidating simple, hand-tuned heuristics and making blind replication of tuning procedures impractical.

Although some technological splits have to be investigated through dedicated test geometries (e.g. Hall bars), some of the gate stack integration, as well as the gate geometry, imply direct testing of quantum devices close to their final operating regime. In this regard, it is necessary to develop a characterization tool for quantum dots in the few-charge regime, robust to important response changes. Moreover, it has to be automated in order to probe the statistical response of a given device geometry, without eventual biases caused by microscopic effects.

The work in this paper focuses on solving the tuning roadblock, specifically the charge tuning step, and was tested on silicon \glspl{qd} fabricated on 300\,mm wafers using industrial grade, \gls{cmos} compatible processing and measurement tools \cite{Bertrand2023,Bedecarrats2021,maurand_cmos_2016}. We developed a pipeline based on \gls{dl} \cite{lecun2015deep} for automatic charge state tuning of silicon \glspl{qd}. We acquired and manually annotated a large dataset of 1015 \gls{csd} spanning nine distinct device designs, multiple process batches, and different wafers (for a complete breakdown of mask/layout/batch/wafer counts and representative examples, see Supplementary Sec.~1). We then trained a U-Net style \gls{cnn} \cite{Ronneberger2015} to segment charge transition lines and evaluated it using five-fold group cross-validation. The trained model produces per-pixel segmentation probability maps that, after postprocessing, locate the first two transition lines and thereby tune the devices to the single charge regime. We tested this framework on real field data with low contrast and noisy CSDs due to experimental cryogenic conditions, achieving an overall tuning success rate of 80.0\%  for locating the single charge regime, with peak per-design performance reaching 88\%. We complement these quantitative results with a detailed failure analysis that identifies common error modes (missed, faint, spurious, and fragmented lines) and propose mitigations. Finally, we outline a roadmap for integrating the developed approach into cryogenic wafer probers, as well as extracting physics-based features from the measured \glspl{csd} using this method. A schematic summary of the experimental setup, representative device geometries, and the offline auto-tuning pipeline is shown in Fig.~\ref{fig:Summary}.

\begin{figure*}[t!h]
    \centering
    \includegraphics[width=\textwidth,keepaspectratio]{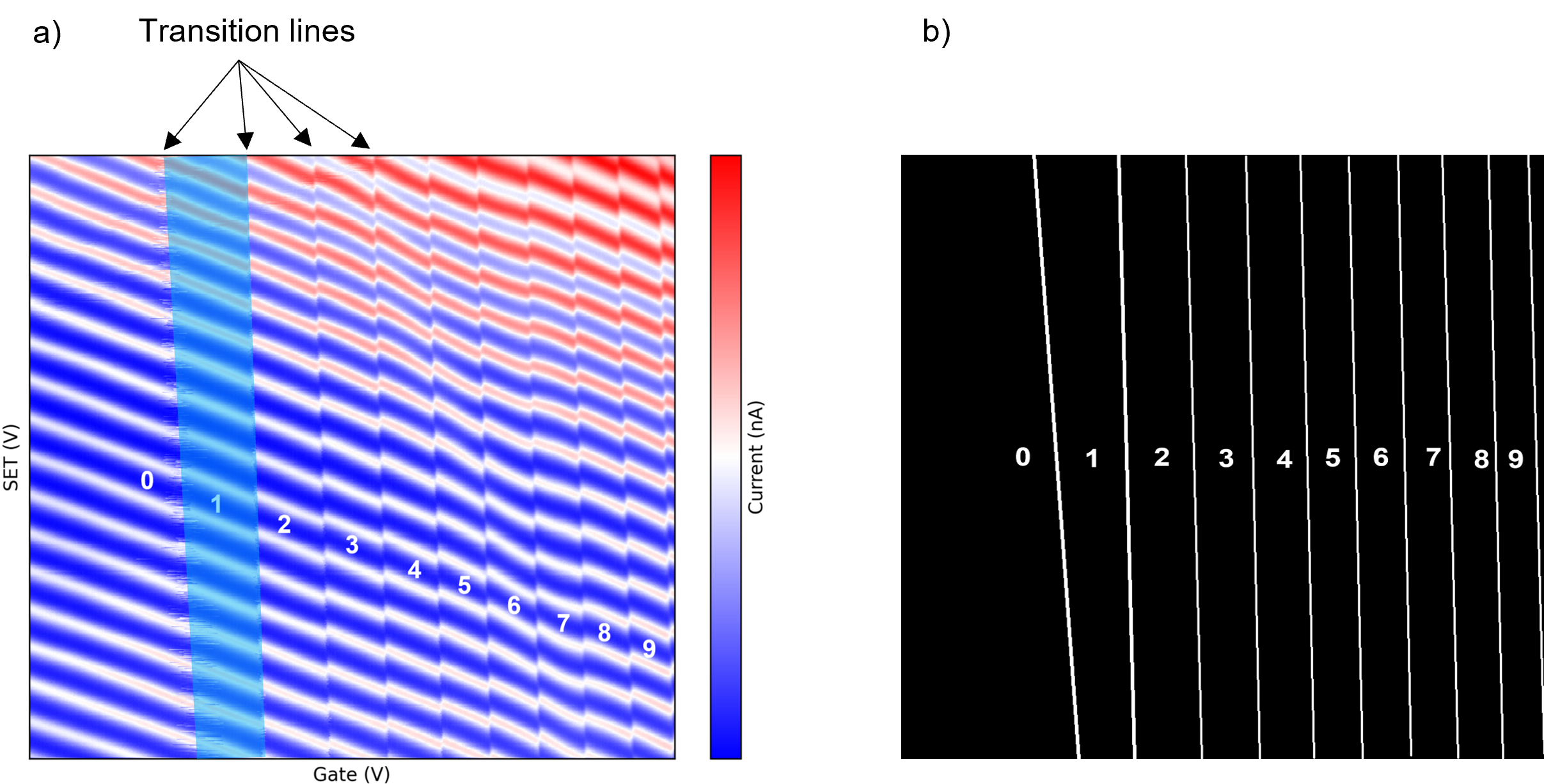}
    \caption{(a) Charge stability diagram: a 2D current map associated to a \gls{qd}. The x and y axes represent the voltage sweep applied to the gates, and the color bar represents the measured current value at every pixel in the map. The numbers represent the charges that are trapped in the dot, separated by charge transition lines. The area highlighted in light blue corresponds to the single-electron regime. (b) Ground truth mask of the \gls{csd} obtained by manually labeling the transition lines.}
    \label{fig:csd_and_mask}
\end{figure*}

\section{Background and Scope}
The tuning process of a gate-defined \gls{qd} is a multi-step process that requires careful consideration before the device can be used to encode a spin-qubit. We refer to \cite{Zwo23} for a detailed description, but here we highlight the main steps. i) First, \textit{bootstrapping} is performed, where the device is cooled down, and its charge sensor initialized. ii) Second, the topology of the device is established through \textit{coarse tuning}, which consists of reaching a known configuration of charge islands with well-defined connectivity between themselves and the reservoirs through tunnel coupling \cite{Ziegler2023, ziegler_toward_2022,liu_automated_2022}. iii) Next, \textit{controllability} is ensured by introducing virtual gates to compensate for capacitive crosstalk between the dots \cite{perron_quantitative_2015}. iv) Then comes the \textit{charge tuning} stage during which we manipulate the charge occupancy of the device by trapping the desired amount of electrons or holes in each \gls{qd} \cite{Baart2016, LapointeMajor2020, Ziegler2023, Durrer2020, Czischek2022, Yon2024,Yon2025,Schuff2024}. Finally, \textit{fine tuning} is performed by adjusting the tunnel coupling between the dots in order to achieve precise control over the qubit's state \cite{teske_machine_2019}. In this paper, we focus exclusively on the \emph{charge tuning} stage, in which the goal is to trap an exact number of charges in each dot.

\subsection{Charge tuning process}
Charge tuning is typically performed by recording a \gls{csd}, which is a two-dimensional map representation of current as a function of gate voltages, as we can see in Fig.~\ref{fig:csd_and_mask}(a). In our case, a \gls{set} serves as the charge sensor, detecting shifts in signal that arise from changes in the surrounding charge state. By sweeping the voltages on both the \gls{qd} and the \gls{set}, a 2D \gls{csd} is obtained, where each pixel corresponds to a measured current value. The Coulomb blockade effect produces a characteristic grid of transition lines, where each line indicates the addition or removal of a single charge. Knowing that an n-type \gls{qd} device is empty at low gate voltages (or high gate voltages for a p-type device), the experimentalist can locate the single charge regime by identifying the first transition line, which represents the zero-to-one charge transition. Therefore, the operator can set the device to the single electron (or hole) regime by applying the gate voltages corresponding to that transition. In practice, this is done by iteratively zooming and re-scanning certain regions of the \gls{csd}, relying on visual heuristics to identify genuine transition lines and distinguish them from spurious dots or line-like artefacts induced by defects, impurities, instrumentation glitches, or measurement noise. While this interactive workflow is tractable for a single large device containing many nominally identical dots (where the same heuristic can be reused), it becomes impractical in a high-throughput cryogenic wafer-prober setting in which many distinct, heterogeneously fabricated small devices must be tuned independently.

\subsection{Limitations of prior approaches}
To address this bottleneck, researchers have pursued two broad automation strategies. On one hand, classical script-based algorithms implement hand-crafted decision rules and can perform well when device geometry and measurement conditions are uniform, but they do not generalize readily to different device layouts, process-induced variability, or unexpected artefacts and typically still require human intervention \cite{Baart2016,LapointeMajor2020,Ziegler2023}. On the other hand, \gls{ml} methods learn tuning-relevant patterns from data and are therefore better suited to handle inter-device heterogeneity. Accordingly, the recent trend favors \gls{ml} approaches \cite{Durrer2020,Czischek2022,Yon2024,Yon2025}. Most prior \gls{ml} work has adopted a patch-based approach in which a fully connected \gls{nn} or a \gls{cnn} classifier has been trained on small regions of the stability diagram to infer if the patch contains a transition line or not, and through an iterative process, a tuning algorithm would attempt to locate the single charge regime by exploring a subregion of the \gls{csd}.

The patch-based approach was originally motivated by its ability to reduce data-acquisition time: instead of measuring a full \gls{csd}, the algorithm inspects a sequence of small subregions in an iterative exploration to locate the desired charge regime. However, this gain in speed comes at the cost of reduced robustness. On one hand, patch-based methods accelerate measurements by focusing only on a subregion of the diagram, but on the other hand, they ignore the global context of the \gls{csd}, making them sensitive to low-frequency noise, discontinuities, and device-to-device variability observable only at larger scales. Moreover, this approach is specialized in finding a specific charge regime, it requires additional iterations and more exploration time if we want to locate an additional regime in the same diagram. Additionally, because it does not reconstruct the full \gls{csd}, the patch method is of limited use for debugging, extracting device physics, or providing feedback to design and fabrication workflows of the devices, it is essentially a local tuning method rather than a characterization tool.

\subsection{Proposed method}
Recent developments in measurement hardware make it possible to routinely acquire a full \gls{csd} at a rate of approximately 1~\si{\micro\second} per pixel \cite{champain2024real}. Taking these advances into account, we propose a wide-diagram semantic segmentation approach that overcomes the limitations of the patch-based methods. Our approach ingests a large span of the \gls{csd} in a single inference pass and locates all transition lines simultaneously, allowing us to identify any desired charge regime without additional measurement and detection steps. In this work, we train a U-Net style \gls{cnn} on a diverse set of manually annotated \glspl{csd}, enabling the model to learn global, physically relevant patterns from a holistic view of the diagram. This makes the model robust to noise and spurious artifacts, reducing misclassifications and improving accuracy. Moreover, the per-pixel predictions produced across the full diagram naturally yield physical parameters such as transition-line slopes, directions, and separations, which can be used for device characterization. Therefore, our proposed method serves dual purposes: (i) charge tuning of \glspl{qd} into any desired regime, and (ii) characterization to provide feedback that helps improve device fabrication and development.

\subsection{Experimental validation}
In order for any tuning algorithm to claim to be an improvement over classical hardcoded methods and to be useful across different devices, it must be tested across the full range of fabrication variability encountered in practice. In this work, we validate our framework on a large and diverse dataset of 1015 experimental \glspl{csd}, measured from two distinct mask designs, fabricated in CEA-Leti's industrial-grade cleanroom facilities across four process batches on seven separate 300-mm wafers, and covering both n- and p-type devices as well as nine different gate pattern geometries of silicon \glspl{qd}. By spanning multiple mask layouts, polarities, batches, wafers, and device geometries, the dataset captures the variety across process fabrication and therefore provides a good benchmark of a model's ability to generalize. This scale and diversity stand in contrast to prior experimental validations, which typically trained and tested on only a few experimental \glspl{csd} or on repeated measurements of the same device, limiting their ability to demonstrate robustness to variations \cite{Yon2025,Yon2024,Czischek2022,Durrer2020}.

\begin{figure*}[t!h]
  \centering
  \includegraphics[width=\textwidth]{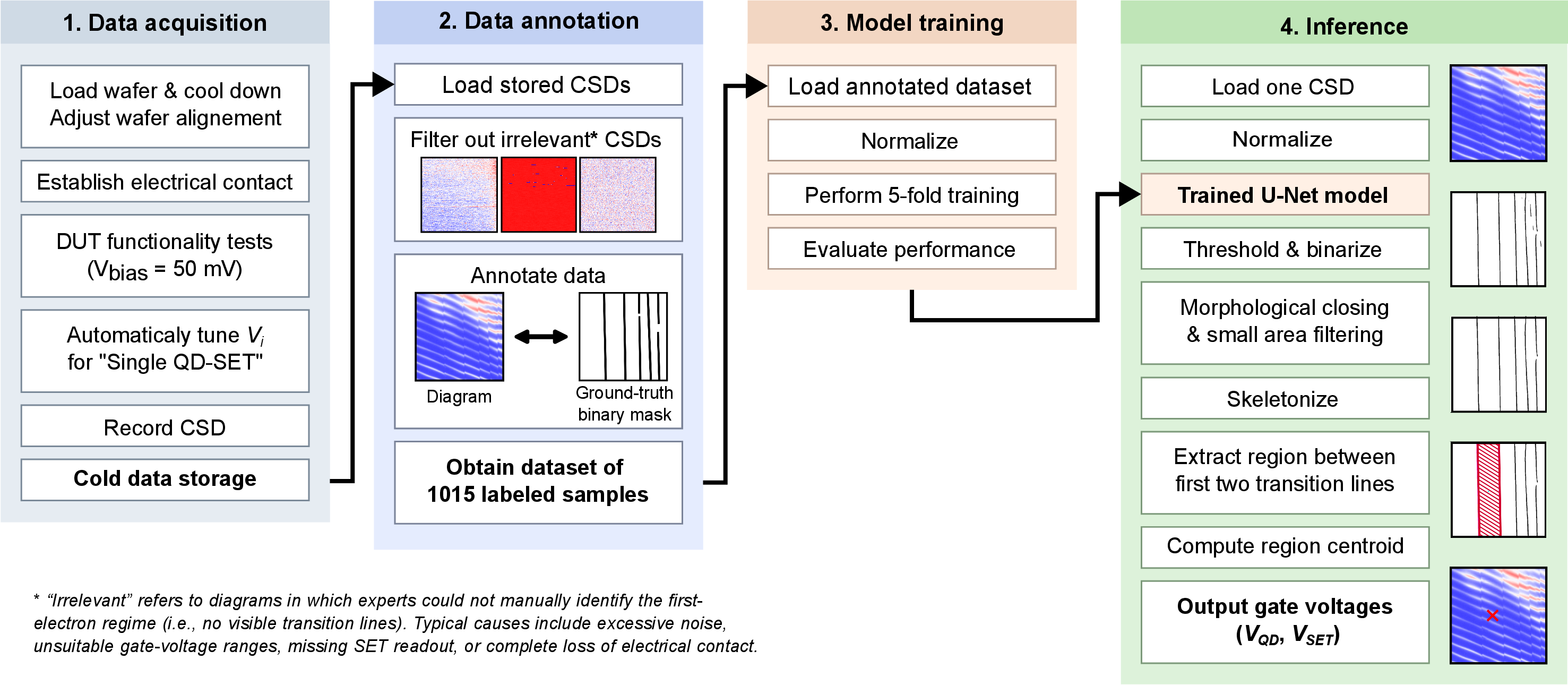}
  \caption{Offline auto-tuning pipeline. From left to right: (1) Cryogenic data acquisition of full \glspl{csd}. (2) Manual annotation and dataset assembly (1015 labeled \glspl{csd}). (3) U-Net model training with five-fold group cross-validation. (4) Inference and postprocessing, (thresholding, morphological closing, dynamic component filtering, skeletonization) to extract the single-charge regime and return gate-voltage targets. Side panels show typical intermediate outputs produced by the pipeline.}
  \label{fig:offline_flow}
\end{figure*}

\section{Methodology}
Our methodology for automatic detection of the voltage space corresponding to electron regimes comprises a multi-stage pipeline for \gls{csd} data acquisition and labeling, model implementation, training, and segmentation predictions. The full flow is summarized in Fig.~\ref{fig:offline_flow}.

\subsection{Data Acquisition and Labeling}
Our framework was validated in an \emph{offline} auto-tuning workflow, i.e., we performed the charge regime detection experiment on static measurements of \glspl{csd}. All devices considered in this study are \gls{set}-\gls{qd} pairs (an \gls{set} used as a charge sensor is coupled with a gate-defined \gls{qd}). We begin by acquiring full \glspl{csd} of various gate-defined silicon \glspl{qd} patterned on 300 mm wafers. The measurements are done using a cryogenic wafer prober \cite{neyens_probing_2024,Contamin2022,zwerver_qubits_2022} which produces raw 2D current maps that serve as the input to our preprocessing and segmentation pipeline (Fig.~\ref{fig:offline_flow}.1). For supervised training, we generated binary ground-truth masks by annotating transition lines on the measured \glspl{csd}. This step was performed over multiple iterations in which the annotator manually traced lines on each diagram using custom software developed by our team to delineate the charge transitions (Fig.~\ref{fig:offline_flow}.2). This was later rasterized to generate a binary mask where black pixels represent the background and white pixels represent the transition lines as seen in Fig.~\ref{fig:csd_and_mask}(b). These masks encode the expert's knowledge and intuition into a ground truth image to be used for the \gls{ml} model to optimize its loss function on. In order to minimize labeling noise and subjective bias, we adopted a multi-annotator review protocol in which every \gls{csd} was labeled by one annotator and subsequently checked by a second. An ambiguity flag was recorded for borderline cases, in which the annotators were not sure about the accuracy of the labeling process, and persistent disagreements were resolved by short consensus reviews. This conservative labeling strategy ensures the training targets reflect experimental practice and quantum devices' expert judgment. The labeled dataset intentionally spans both high-quality, easily annotated diagrams and low-quality or defective devices, ensuring the model is trained and validated across the full range of practical measurement conditions.

\subsection{Model Architecture and Training}
The core component of this framework is the \gls{ml} model we developed, which consists of a \gls{cnn} built in a U-Net architecture \cite{Ronneberger2015} with MobileNetV2 as its lightweight encoder \cite{Sandler2018}, pretrained on ImageNet for faster convergence and improved feature generalization \cite{deng2009imagenet}. The choice of MobileNetV2 offers a good trade-off between accuracy and computational efficiency. Its use of depthwise separable convolutions and inverted residuals greatly reduces computational load and memory footprint, which is advantageous for deployment in resource-constrained environments, making it suitable for both local inference on laboratory workstations and potential in-situ integration within cryogenic or embedded control hardware. Precise layer-by-layer counts, the total parameter breakdown, and model topology are given in Supplementary Sec.~3.

This is a pixel-wise segmentation task, where for every pixel in the input \gls{csd} the model predicts the probability that it belongs to a transition line or to the background. Training was performed in a supervised manner using \glspl{csd} and their corresponding ground-truth masks, with the \textit{Dice} loss as the optimization objective \cite{Milletari16}.
The \textit{Dice} loss \(\mathcal{L}_{DICE}\) is defined as
\begin{equation}
\label{eq:dice_loss}
\mathcal{L}_{\it{Dice}} \;=\; 1 - \frac{2\sum_{i} p_i g_i}{\sum_{i} p_i + \sum_{i} g_i + \varepsilon},
\end{equation}
where \(p_i\in[0,1]\) is the model's predicted probability for pixel \(i\), \(g_i\in\{0,1\}\) is the ground-truth binary label, and \(\varepsilon>0\) is a small constant for numerical stability.

We perform the offline auto-tuning experiment using five-fold group cross-validation on the full dataset. For each fold, the model is trained on the union of four folds and evaluated on the remaining held-out test fold. This procedure yields five independent test results, which are then aggregated to report the overall performance. To prevent data leakage, all \glspl{csd} originating from the same physical device (i.e., same design, die, wafer, etc.) are assigned to a single fold, ensuring that no device appears simultaneously in both the training and test sets.

\begin{figure*}[t!h]
  \centering
  \begin{tabular}{ccc}
    \includegraphics[width=0.32\textwidth,keepaspectratio]{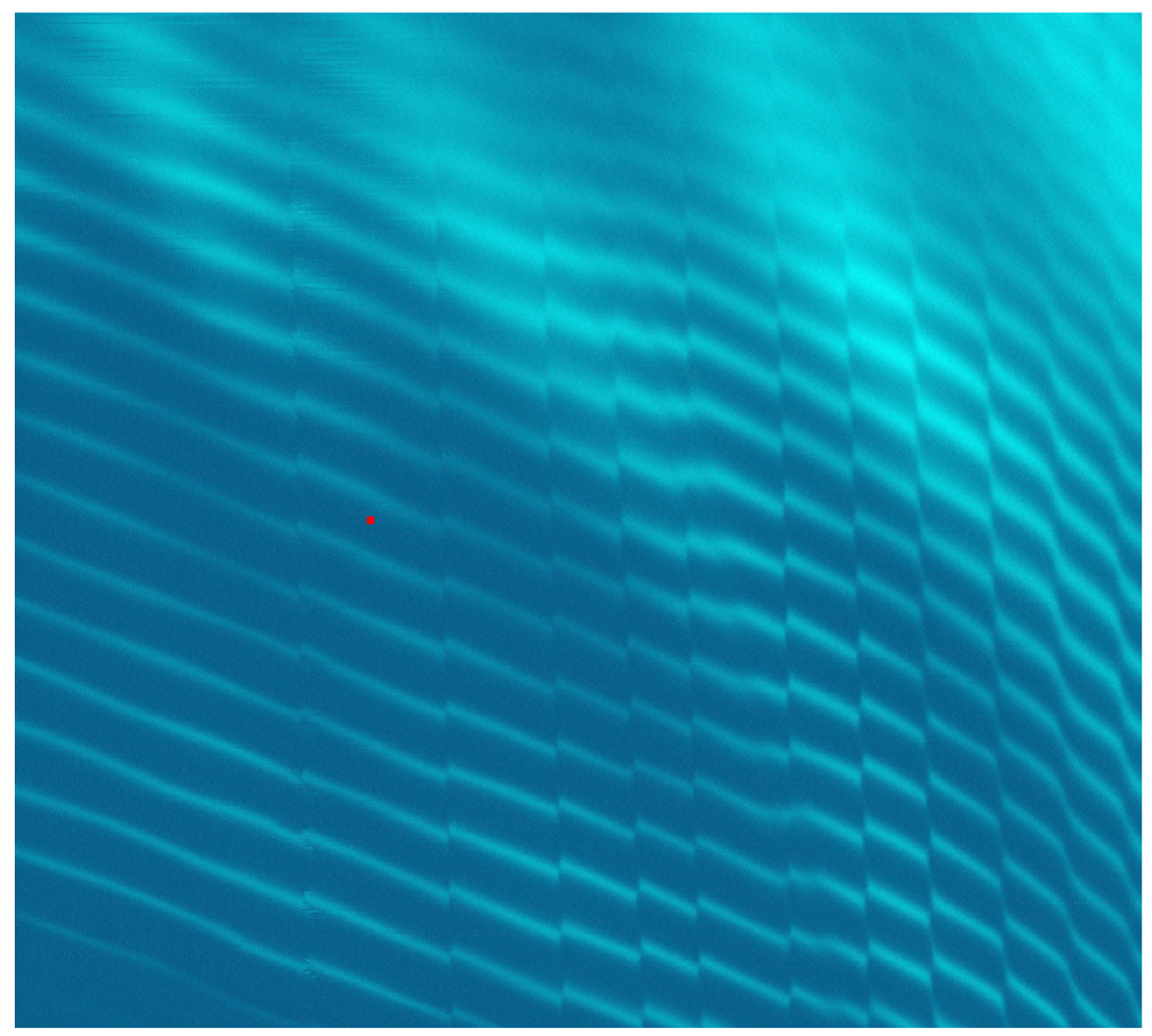} &
    \includegraphics[width=0.32\textwidth,keepaspectratio]{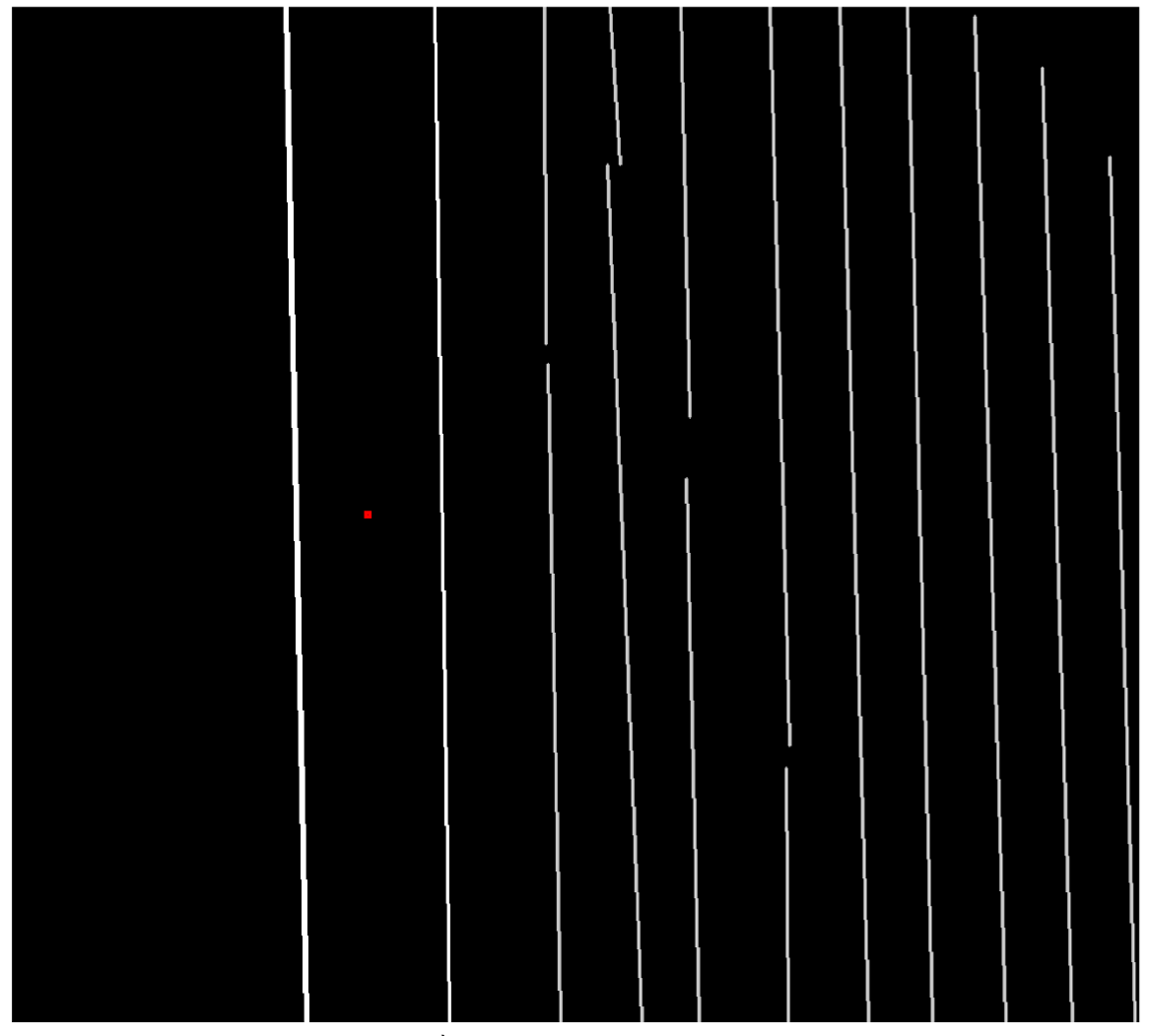} &
    \includegraphics[width=0.32\textwidth,keepaspectratio]{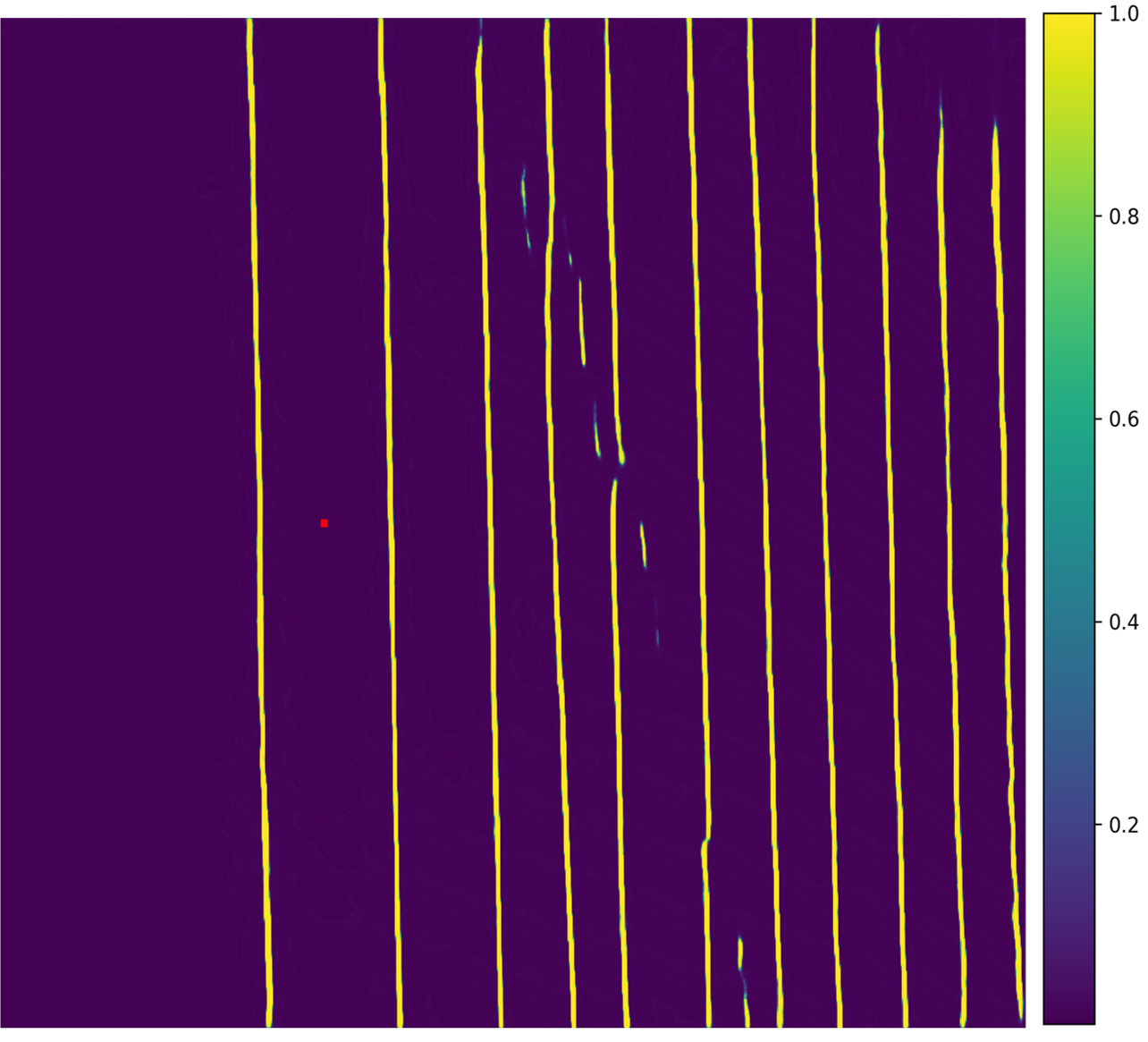} \\
    \small (a) Input Image & \small (b) Ground Truth Mask & \small (c) Prediction Probability Map \\[8pt]
    \includegraphics[width=0.32\textwidth,keepaspectratio]{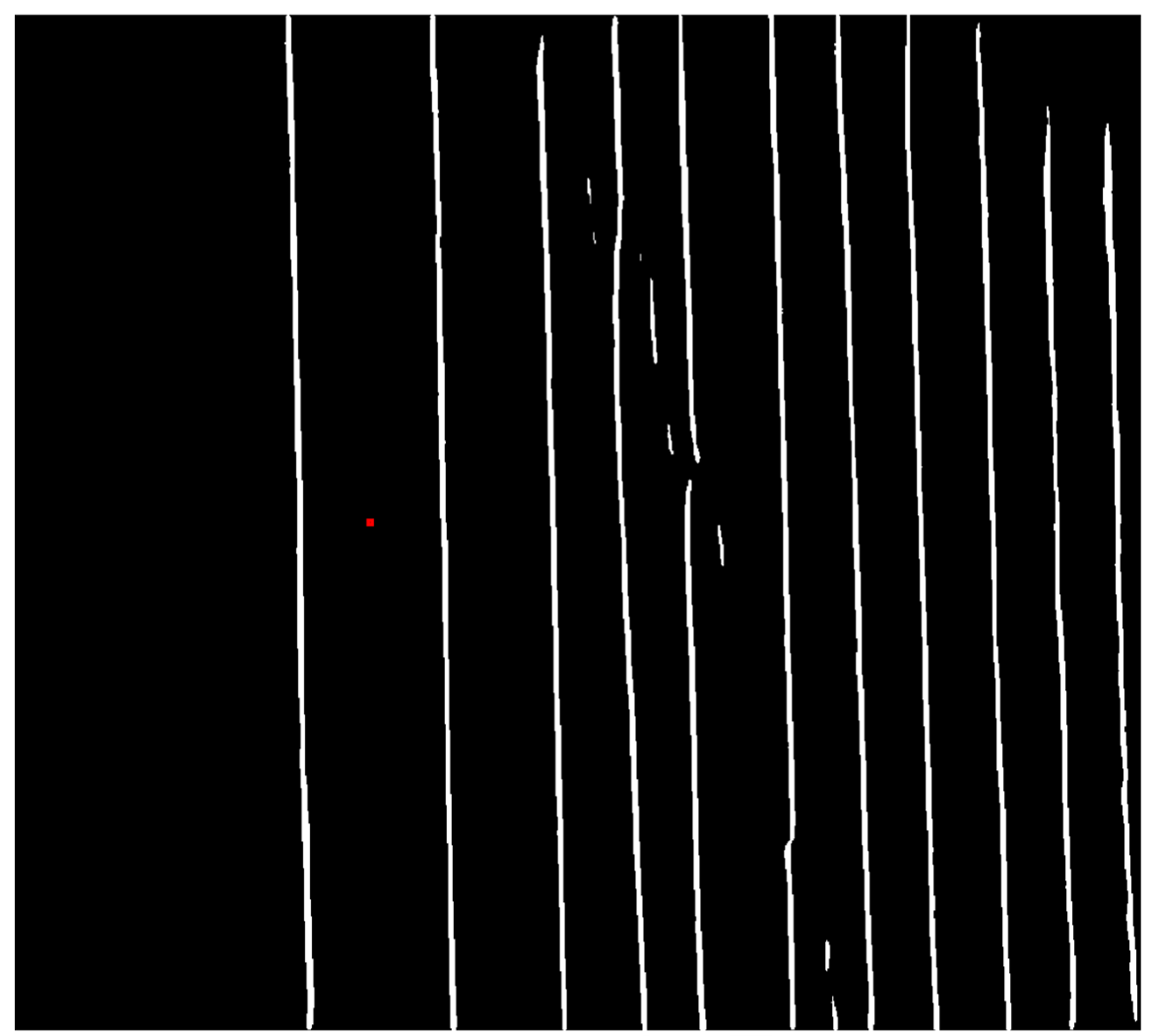} &
    \includegraphics[width=0.32\textwidth,keepaspectratio]{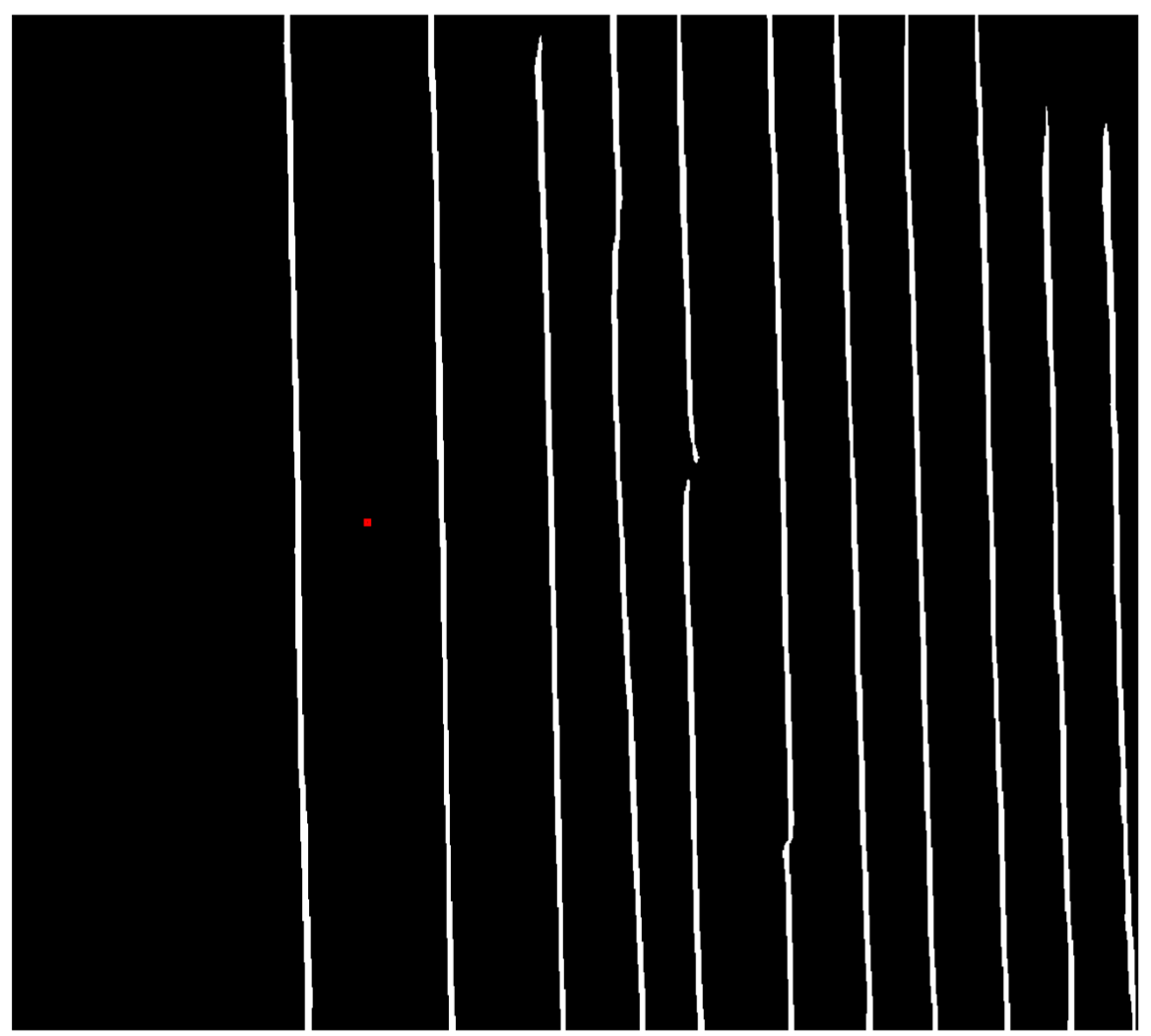} &
    \includegraphics[width=0.32\textwidth,keepaspectratio]{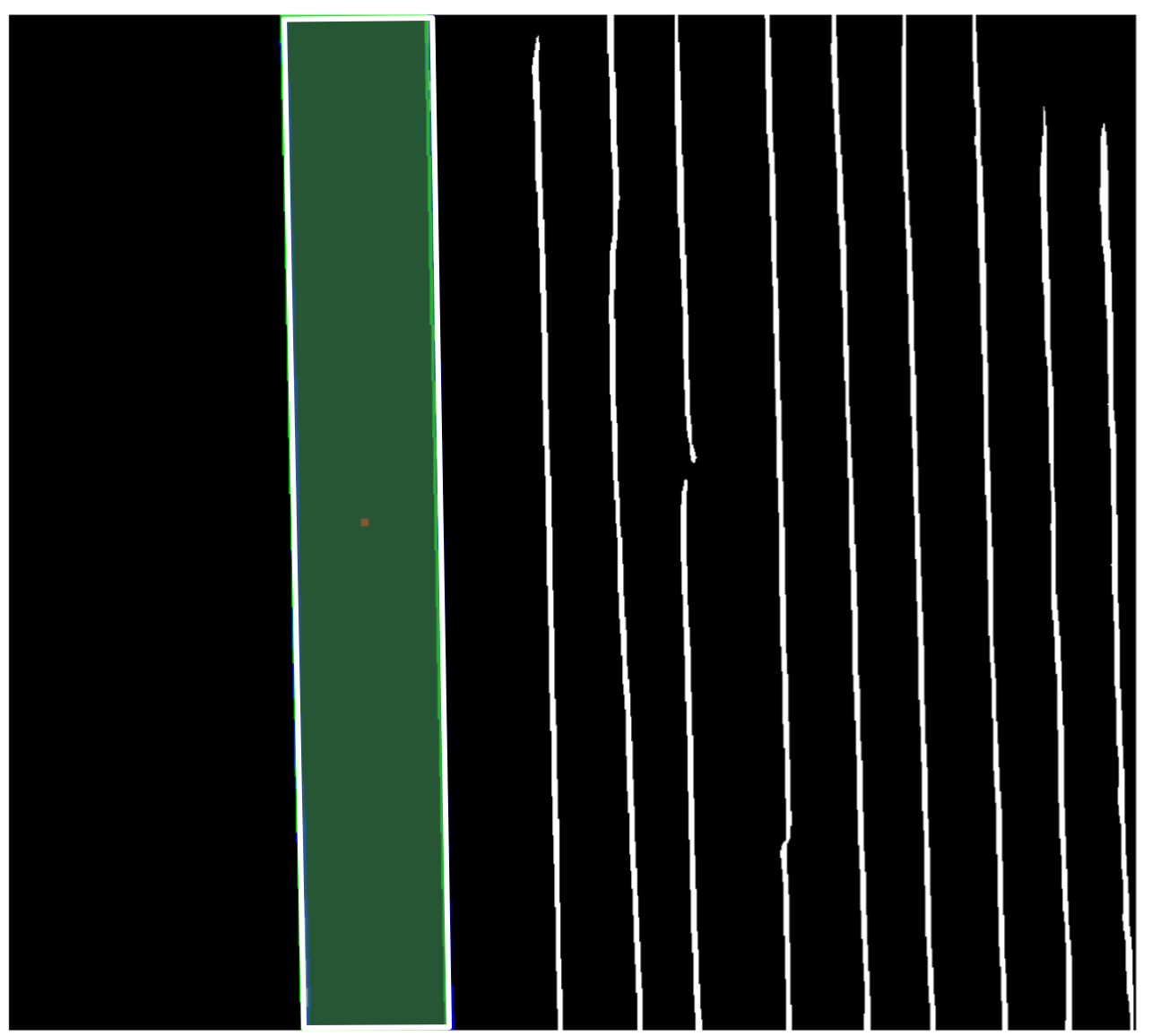} \\
    \small (d) Binarized Map & \small (e) Morphology + Filtering & \small (f) Single Charge Regime
  \end{tabular}
  \caption{Plot showing inference and the main postprocessing steps. (a) Input \gls{csd} as an image. (b) Expert annotated ground-truth transition-line mask used for training. (c) Raw model output: per-pixel probability map (colormap indicates model confidence between $[0,1]$ that the specific pixel belongs to a transition line). (d) Binary map obtained by thresholding the probability map at $0.75$. (e) Morphological closing (to bridge small gaps) followed by dynamic area filtering / connected-component area thresholding (to remove small spurious detections) that connects broken segments and suppresses parasitic features. (f) Final cleaned binary mask from which the first two transition lines are extracted, and the center of mass of the single charge (1e$^-$) regime is computed (highlighted here in green). This center gives the gate-voltage target proposed by the pipeline for charge tuning.}

  \label{fig:postprocessing-steps}
\end{figure*}

\subsection{Prediction and Post-Processing}
\label{ssec:postprocessing}

We adopt a task-level \emph{tuning success} metric as the primary evaluation: a detection is successful when the predicted single-charge center falls inside the annotated single-charge region. This choice reflects the pipeline's operational objective of returning gate voltage targets for device tuning, and was therefore used to validate our method. Segmentation outputs were visually and functionally inspected to confirm they provide the geometric fidelity required for downstream localization.

The network produces a per-pixel probability map that is converted to a binary line mask and refined by a short deterministic post-processing chain (illustrated in Fig.~\ref{fig:postprocessing-steps}). First, the probability map is thresholded at \(\tau=0.75\) based on empirical validation. To connect fragmented line segments, we apply morphological closing with a vertically biased structuring element to bridge short gaps along predominantly vertical transition lines present in our stability diagrams. Small spurious detections are then removed by a dynamic area filter. It treats the areas of connected components as a distribution and discards components whose areas fall under a diagram-dependent threshold. After filtering, the predicted transition lines in the binary mask are sorted based on their horizontal coordinates. The two most extreme components (lowest or highest, depending on device polarity) are selected as the bounding transitions of the single-charge region. The centroid of the polygonal region between these two lines is taken as the predicted single-charge center, and the pixels inside of that region are converted to physical gate voltages via the linear pixel to voltage mapping encoded in the scan metadata. Exact parameter values and additional details on the postprocessing and auto-tuning algorithm are reported in the Supplementary Materials Sec.~4.

\begin{figure*}[t!]
  \centering
  \begin{tabular}{ccc}
    \small (a) Input CSD & \small (b) Ground-truth mask & \small (c) Prediction \\[4pt]
    \includegraphics[width=0.32\textwidth,keepaspectratio]{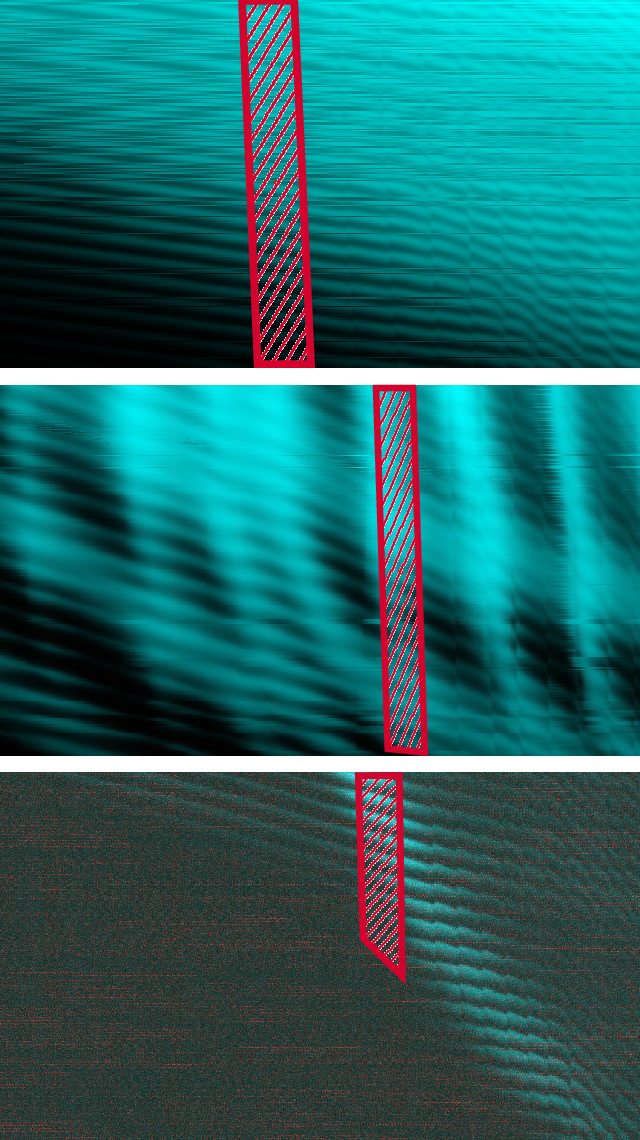} &
    \includegraphics[width=0.32\textwidth,keepaspectratio]{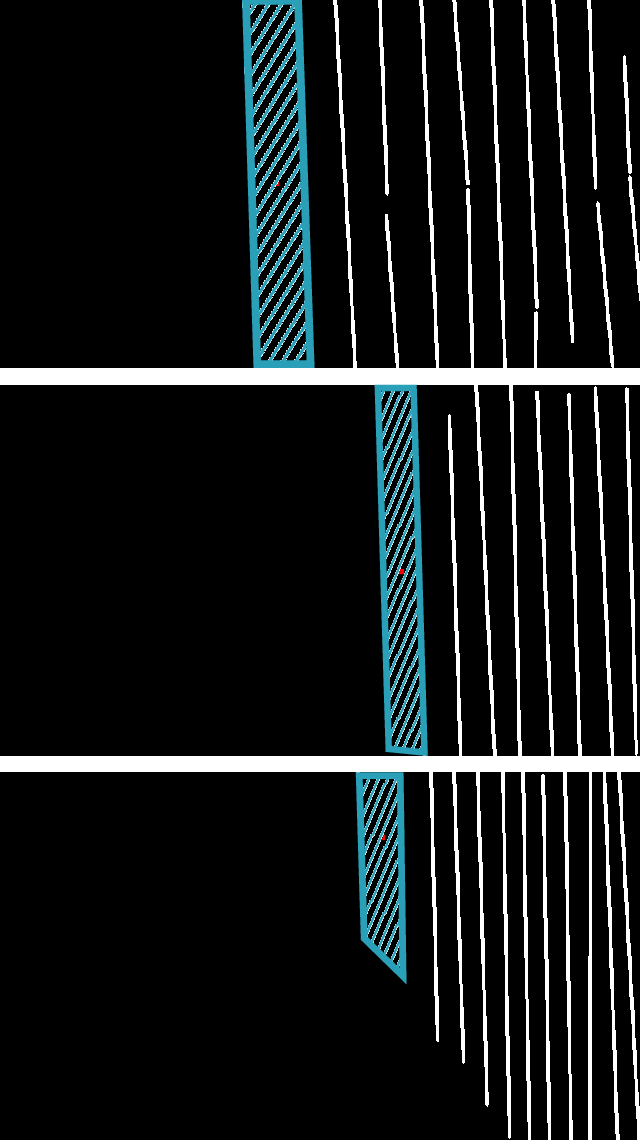} &
    \includegraphics[width=0.32\textwidth,keepaspectratio]{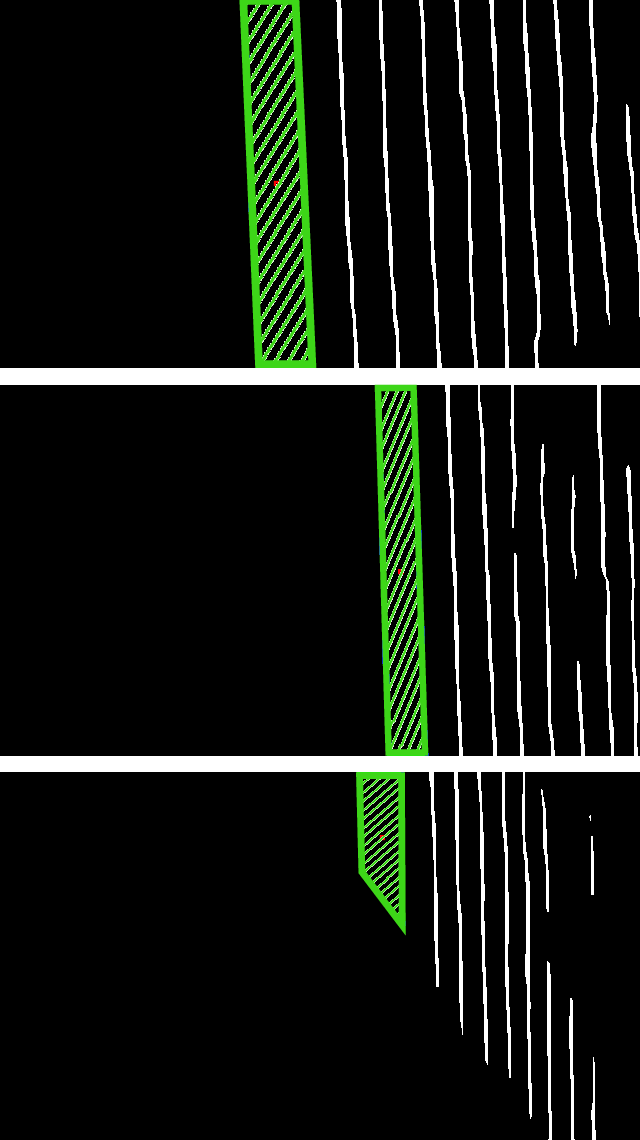} \\
  \end{tabular}
  \caption{\textbf{Representative offline auto-tuning inference.} (a) Measured \gls{csd} (input). (b) Expert-annotated ground truth transition line binary mask. (c) Model prediction after binarization and post-processing; the detected single charge regime center is marked with a red square. The single charge regime is highlighted with red, blue, and green for (a), (b), and (c), respectively. The example highlights how the pipeline extracts the relevant transition lines and proposes gate-voltage coordinates for tuning.}
  \label{fig:results}
\end{figure*}

\section{Results}
\label{sec:results}

\subsection{Single Charge Regime Detection}

We evaluate the offline auto-tuning pipeline using a five-fold cross-validation on the full dataset split by device design. For clarity and to protect confidential design details, we refer to device geometries and mask layouts using anonymized identifiers throughout the manuscript. The nine distinct gate geometries are labeled 'Design A' through 'Design I', and the two mask layouts are labeled 'Mask I' and 'Mask II'.  More details can be found in the Supplementary Materials Sec.~1. Below, we first illustrate a representative inference result and make explicit our definition of a successful tuning. We then report aggregated, per design task-level performance. Figure~\ref{fig:results} presents a typical example of the model's output on a measured \gls{csd}. The detected single charge regime is highlighted with a green overlay, with a red square in its center.

Table~\ref{tab:device_aggregate} summarizes the per-design tuning success across the full dataset. All results reported in that table are aggregated over the five cross-validation test folds. The overall offline tuning success across all 1015 diagrams is 80.0\% (812/1015), while per-design performance ranges from 61\% to 88\%. This spread in per-design performance largely stems from systematic differences in device quality and fabrication runs. It can also be explained by data quality: designs with clear, high \gls{snr} transition lines tend to yield higher success rates, whereas designs with faint, stochastic, or blank regions produce lower scores. This result sets a new benchmark for future work, as it is the first time that a charge tuning algorithm was tested on a large number of \glspl{csd} measured from a wide variety of devices.

At the mask level, Mask I achieves 589/695 (84.7\%) and Mask II achieves 223/320 (69.7\%). This performance gap can be mainly attributed to two factors. First, Mask I is approximately twice as represented in the training data, so the model has learned more of its characteristic patterns. Second, Mask I was designed with enhanced charge sensing and therefore yields well-defined \glspl{csd}, whereas Mask II contains wafers with poor charge sensing, leading to lower-quality \glspl{csd}. The mask-level comparison should therefore be interpreted in light of both training set imbalance and intrinsic design differences.

A closer examination shows that the low performance for Design G is attributable mainly to spurious lines caused by measurement noise rather than a single bad wafer or systematic fabrication defect, while Design H contains many low-quality \glspl{csd} with poorly defined transitions. These observations motivate targeted mitigation such as additional measurements for under-sampled designs, a lightweight quality-check classifier to reject very low-\gls{snr} maps, and online confirmation scans.

\begin{table}[h!t]
  \centering
  \caption{Per-design aggregate tuning success computed over five folds. ``Success (\%)'' is the fraction of \glspl{csd} for which the predicted single charge regime center falls within the annotated charge region. The third column lists the raw successful detections and total diagrams per design.}
  \label{tab:device_aggregate}
  \begin{tabular}{c c c c}
  \toprule
  Mask & Design & Success/Total \glspl{csd} & Success (\%) \\
  \midrule
  I  & Design A & 70/83   & 84\% \\
  I  & Design B & 52/61   & 85\% \\
  I  & Design C & 64/81   & 79\% \\
  I  & Design D & 130/147 & 88\% \\
  I  & Design E & 122/138 & 88\% \\
  I  & Design F & 121/142 & 85\% \\
  I  & Design G & 30/43   & 70\% \\
  II & Design H & 76/124  & 61\% \\
  II & Design I & 147/196 & 75\% \\
  \midrule
  \multicolumn{2}{l}{\textbf{Overall Success}} & \textbf{812/1015} & \textbf{80.0\%} \\
  \bottomrule
  \end{tabular}
\end{table}

\subsection{Failure analysis}
\begin{figure*}
    \centering
    \includegraphics[width=1\linewidth]{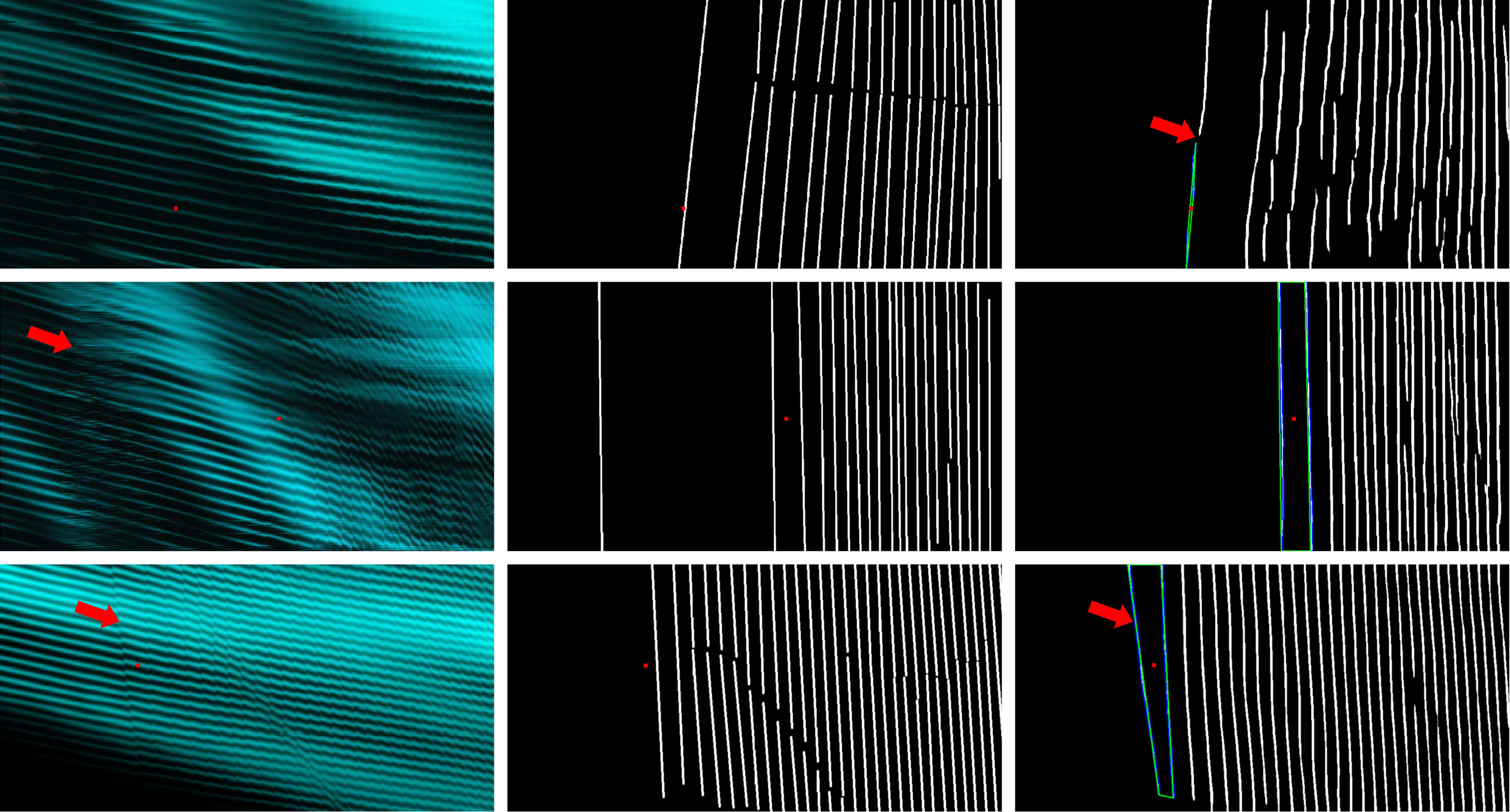}
    \caption{Representative failure-mode examples. Column 1: a stochastic/faint transition that intermittently disappears in the prediction; Column 2: a spurious line-like measurement artefact that can be misclassified as a transition; Column 3: a true transition whose predicted line is fragmented. Each column shows (top) the input \gls{csd} with a red arrow indicating the problematic feature, (middle) the hand-annotated ground-truth transition-line mask, and (bottom) the model's binary prediction and extracted transition contours (blue/green) with the red dot marking the computed single charge target.}
    \label{fig:failure_modes}
\end{figure*}

We conducted a systematic failure analysis to understand the dominant error modes that limit the tuning success. A summary of the observed failure types and the proposed mitigations is given in Table~\ref{tab:failure-modes-mitigation}. The most frequent causes of failure are defective devices that cause low-quality \glspl{csd} with very low \gls{snr} or large blank regions, faint or stochastic transition lines that elude reliable detection, spurious line-like artefacts introduced by measurement or instrumentation, fragmented predictions where true transitions are broken into multiple segments, and occasional annotation ambiguity between borderline features. These failure modes are not mutually exclusive and often appear together in the most challenging examples. Fig.~\ref{fig:failure_modes} contains some examples of failure modes that were commonly observed during the experiment.

Table~\ref{tab:failure-modes-mitigation} summarizes targeted mitigations derived from the failure-mode analysis. Practically, the most impactful near-term steps are (i) instituting an automated quality-check classifier to exclude clearly unusable \glspl{csd} before attempting tuning, (ii) adding online confirmation scans when the model flags ambiguous or low-confidence regimes, and (iii) strengthening post-processing to better merge fragmented segments and reject spurious frequency-orientation outliers.

We emphasize the following when interpreting the reported performance: the dataset used for evaluation deliberately retained defective and low-quality devices rather than pre-filtering them out. As a result, the reported 80\% aggregate auto-tuning success should be interpreted as a conservative measure that reflects two factors: the model's ability to predict the single-charge regime \emph{and} the intrinsic fabrication and measurement quality of the devices. Removing clearly unusable devices or applying a quality pre-filter prior to tuning would increase the effective success rate. In practice, these devices would not be used for qubit implementation, and therefore, we do not need to tune them. By implementing a classifier that would flag defective devices before the tuning step, we would achieve a higher tuning success rate that is more representative of the downstream task of qubit implementation. Implementing the recommended mitigations, particularly pre-filtering and online confirmation scans, constitutes a direct and practical route to improving auto-tuning reliability.

 It is important to emphasise the measurement context: unlike many prior studies that tune a carefully selected device (cooled and hand-checked in a dedicated cryostat before auto-tuning is applied), our pipeline is designed for automated wafer probing, where hundreds of nominally identical devices are measured sequentially with minimal per-device manual intervention. Automatic wafer probing, therefore, exposes the full spectrum of as-fabricated behavior, rather than only well-behaved devices. For this reason, we intentionally retained low-quality and defective \glspl{csd} in the evaluation. Even under these demanding conditions, an 80\% automatic detection rate of the single charge regime already enables reliable parameter extraction for the majority of devices in an automatic wafer-prober setting and is sufficient to deliver useful, large-scale statistics back to fabrication and design teams.

\begin{table*}[b!ht]
  \centering
  \small
  \setlength{\extrarowheight}{4pt}
  \begin{tabular}{lll}
    \toprule
    \parbox[t]{3cm}{\raggedright\textbf{Failure Mode}} &
    \parbox[t]{5.5cm}{\raggedright\textbf{Associated Causes}} &
    \parbox[t]{7cm}{\raggedright\textbf{Failure Mitigation}} \\[6pt]
    \midrule
    \parbox[t]{3cm}{\raggedright\textbf{Bad Quality \gls{csd}}} &
    \parbox[t]{5.5cm}{\raggedright Defective device / Noisy measurements\\ Large regions with no signal (low \gls{snr})} &
    \parbox[t]{7cm}{\raggedright Train a \emph{good vs.\ bad} \gls{csd} classifier\\ Re-measure noisy \glspl{csd}} \\[14pt]
    \parbox[t]{3cm}{\raggedright\textbf{Missed Lines}} &
    \parbox[t]{5.5cm}{\raggedright Stochastic lines\\ Faint lines} &
    \parbox[t]{7cm}{\raggedright Add contrast / gradient channels as inputs\\ Expand / relabel dataset for stochastic lines} \\[14pt]
    \parbox[t]{3cm}{\raggedright\textbf{Spurious Lines}} &
    \parbox[t]{5.5cm}{\raggedright Noise-induced false positives\\ Instrumentation or measurement artefacts} &
    \parbox[t]{7cm}{\raggedright Confirm during online tuning\\ Use FFT to spot lines with different slopes and spatial frequencies} \\[14pt]
    \parbox[t]{3cm}{\raggedright\textbf{Fragmented Lines}} &
    \parbox[t]{5.5cm}{\raggedright Low contrast leading to broken lines\\ Interrupted transitions} &
    \parbox[t]{7cm}{\raggedright Use row-sweep grouping to merge fragmented segments} \\[14pt]
    \parbox[t]{3cm}{\raggedright\textbf{Ambiguous Labeling}} &
    \parbox[t]{5.5cm}{\raggedright Human annotation inconsistencies} &
    \parbox[t]{7cm}{\raggedright Multi-annotator\\ Zoom / re-scan during online tuning} \\
    \bottomrule
  \end{tabular}
  \caption{Summary of common failure modes observed in our stability diagrams, their associated causes, and suggested mitigations.}
  \label{tab:failure-modes-mitigation}
\end{table*}

\section{Discussion and Future Work}
This work contributes three practical advances. First, a scalable full-image segmentation approach that leverages global stability diagrams rather than local patch-based decisions, improving robustness to local noise and ambiguous features. Second, we demonstrate that a manually annotated dataset covering nine device geometries and several wafers and fabrication runs provides enough diversity for the model to generalize across devices. Third, an end-to-end offline auto-tuning pipeline from preprocessing and annotation through inference and parametric postprocessing that yields direct voltage for automated gate updates. We also identify key limitations and directions for improvement. The reported aggregate success is conservative because the evaluation deliberately retained defective devices unfit for qubit implementation rather than pre-filtering them. Performance is sensitive to measurement quality and device-specific characteristics: faint, stochastic, or fragmented lines and spurious artefacts remain the dominant failure modes for which mitigation methods have been suggested.

An important and distinct opportunity enabled by wide-diagram segmentation is physics-based feature extraction. From the predicted binary line maps, we can reliably extract geometric descriptors (line slopes, spacings, and first transition intersection) and map these directly to device-relevant quantities such as transition voltages, inter-line separations (\(\Delta V\)), and approximate lever arms or capacitive couplings. The extracted features facilitate multiple post-processing and analysis tasks, such as providing actionable feedback to fabrication and design teams about systematic variations across wafers. Towards this goal, we have implemented a compact extraction pipeline to validate the physical fidelity of the derived quantities and to prepare these metrics for future studies on the fabrication variation across different devices. We plan to extend this analysis to quantify correlations between fabrication parameters and the extracted features, thereby helping to close the loop from measurement back to process improvement. Previously, this type of feature extraction was performed manually or via simulation on a small number of diagrams, our wide-diagram framework automates the same steps across thousands of \glspl{csd}, substantially reducing the manual workload for experimentalists and further illustrating the practical advantage of a wide-diagram (rather than patch-based) approach. Importantly, the offline mode of our approach is itself highly valuable because the pipeline operates on stored \glspl{csd}, it can be applied retrospectively on archived datasets to extract the same device metrics and to perform the necessary analysis.

Looking forward, the most immediate and impactful next step is closing the loop \textit{in situ} by integrating the model into the control software of a cryogenic wafer prober. Online deployment enables confidence-driven re-measurement (automatic zooms and repeat scans), real-time rejection of spurious detections, and direct measurement of practical gains such as reduced operator effort and per-device tuning time. Achieving this requires engineering work on model compression, low-latency on-edge inference runtimes, and a lightweight decision module that triggers targeted sub-scans when model confidence is low. A practical complement to online integration is a simple pre-filter that removes clearly unusable \glspl{csd} from the tuning pool. When combined with online confirmation scans, this two-stage strategy (filter $\rightarrow$ infer $\rightarrow$ confirm) offers a low-cost path to substantially higher real-world reliability than the offline numbers alone suggest.

 \section{Conclusion}
We presented a wide-diagram segmentation pipeline for automatic charge state tuning of gate defined silicon \glspl{qd}. The method is validated on a large, heterogeneous dataset of 1015 \glspl{csd} measured from devices patterned on 300\,mm wafers with industrial grade, \gls{cmos}-compatible processing and measurement tools. Using a U-Net style \gls{cnn} with a MobileNetV2 encoder, trained with group five fold cross-validation, the pipeline achieves an overall offline tuning success of 80.0\% in locating the single charge regime (812/1015), with peak device-level performance of 88\% on the best performing designs. Beyond tuning, the wide-diagram segmentation produces line maps that enable the extraction of physically meaningful quantities, providing useful feedback to fabrication and device engineering teams.

In summary, this work demonstrates that wide-diagram, \gls{dl} based segmentation is a practical and informative route to automate charge tuning for silicon \glspl{qd}. By combining complementary engineering and algorithmic improvements such as pre-filtering and confidence-driven online confirmation scans, the approach can be matured into a robust component of scalable qubit calibration workflows, helping bridge the gap between experimental quantum hardware and the automated control required for large-scale quantum processors.

\section*{Acknowledgements}
CEA-Leti is a member of the Carnot institute network. This work was supported by the French ANR via Carnot Institute funding. We thank the measurement and fabrication teams at CEA-Leti for providing the devices and measurement infrastructure.

\section*{Conflicts of interest}
The authors declare no conflicts of interest.

\section*{Author contributions}

\noindent \textbf{Peter Samaha}. Designed and implemented the overall pipeline and model architecture. Performed model training and evaluation. Developed data preprocessing and postprocessing routines. Carried out offline experiments and analysis. Wrote the manuscript.

\noindent \textbf{Amine Torki}. Contributed with dataset preparation and engineering infrastructure. Led data annotation and developed the labeling tool. Prepared figures and visual materials.

\noindent \textbf{Ysaline Renaud}. Participated in manual data annotation and quality control of the labeled dataset.

\noindent \textbf{Sam Fiette}. Performed device measurements and acquired \glspl{csd}. Contributed to manual labeling and dataset curation.

\noindent \textbf{Emmanuel Chanrion}. Performed device measurements and acquired \glspl{csd}. Contributed to manual labeling and dataset curation.

\noindent \textbf{Pierre-André Mortemousque}. Acquired funding. Performed device measurements and acquired the \glspl{csd}. Participated in manual annotation review. Contributed to experimental design and data interpretation. Co-supervised the project.

\noindent \textbf{Yann Beilliard}. Acquired funding. Supervised and helped lead the research effort. Provided scientific guidance on experiment design, analysis, and manuscript preparation.

\clearpage
\newpage
\onecolumngrid 

\begin{center}
  {\Large\textbf{Supplementary Material}}\\[0.5ex]
  \textit{Automatic Charge State Tuning of 300 mm FDSOI Quantum Dots Using Neural Network Segmentation of Charge Stability Diagram}
\end{center}

\noindent This supplement provides details on the dataset, labeling workflow, preprocessing, model architecture, training protocol, postprocessing and the exact steps used to convert predicted line masks into gate-voltage recommendations for the single-charge regime.

\section{Dataset}
We collected a heterogeneous dataset of \textbf{1015} charge stability diagrams (CSDs) measured using a cryogenic wafer prober from devices fabricated in CEA-Leti cleanrooms. The dataset spans two mask layouts, nine device geometries, four process batches and seven wafers, and covers both n- and p-type devices. This breadth is purposefully large to test model generalization across realistic process variability.

\begin{figure}[h!]
    \centering
    \includegraphics[width=1\linewidth]{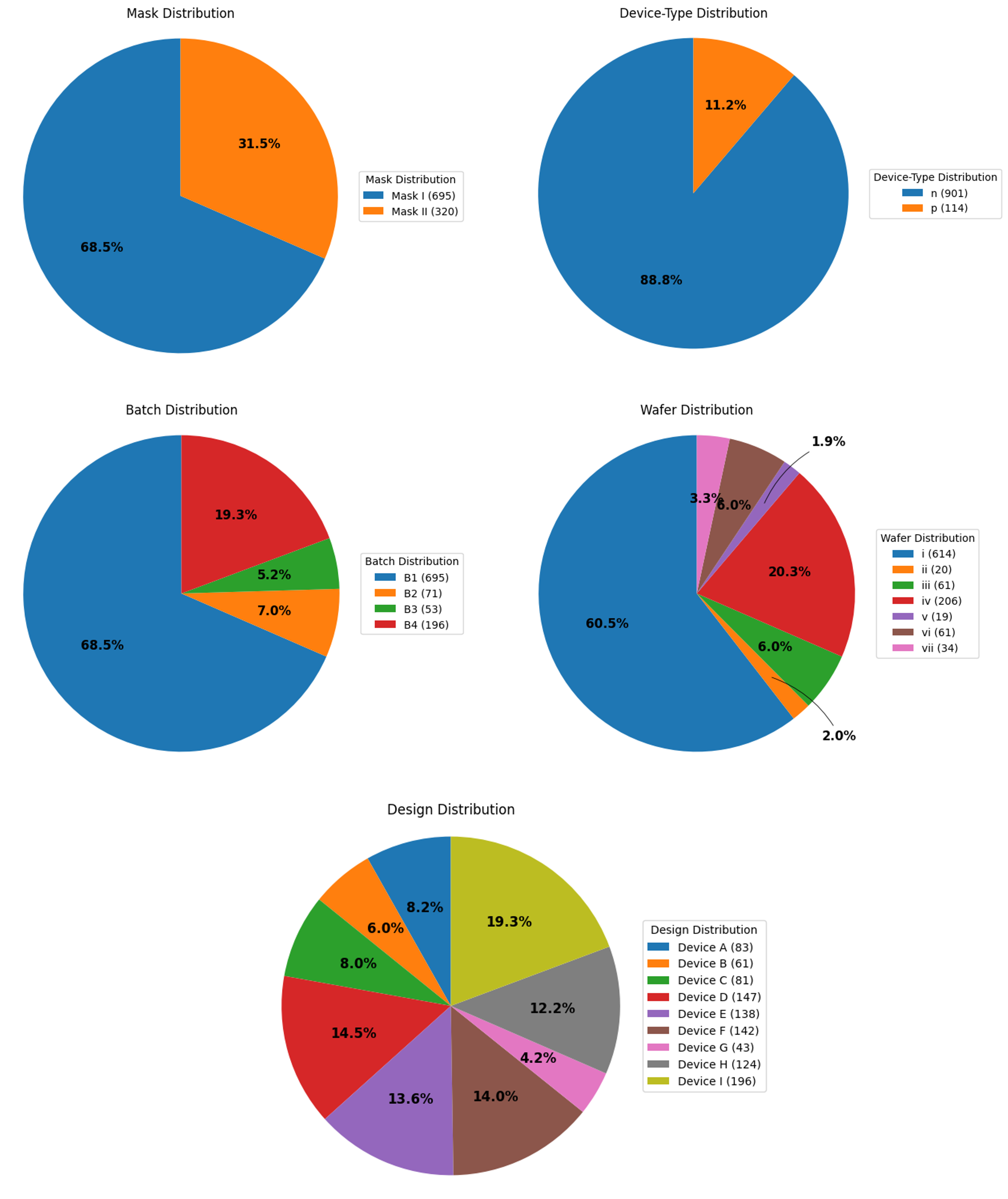}
    \caption{Distribution of our CSDs data across mask designs, device polarity, batches, wafers, and device design, respectively, from top left to bottom right.}
    \label{fig:piechart}
\end{figure}

To illustrate the full hierarchy (mask $\rightarrow$  polarity type $\rightarrow$  batch $\rightarrow$  wafer $\rightarrow$  design), see Table~\ref{tab:hierarchy_summary} below. We anonymized the names for confidentiality.

\begin{table}[ht]
  \centering
  \small
  \caption{Hierarchical breakdown of CSD counts across masks, polarity, batches, wafers, and specific device designs.}
  \label{tab:hierarchy_summary}
  \begin{tabular}{|l|l|l|l|l|c|}
    \hline
    \textbf{Mask} & \textbf{Polarity} & \textbf{Batch} & \textbf{Wafer} & \textbf{Design} & \textbf{Count} \\
    \hline
    \multirow{9}{*}{\textbf{Mask I}}
      & \multirow{9}{*}{n-type} & \multirow{6}{*}{B1} & \multirow{6}{*}{i} & Design A & 83 \\
    \cline{5-6}
      & & & & Design B & 61 \\
    \cline{5-6}
      & & & & Design C & 81 \\
    \cline{5-6}
      & & & & Design D & 147 \\
    \cline{5-6}
      & & & & Design E & 138 \\
    \cline{5-6}
      & & & & Design F & 104 \\
    \cline{3-6}
      & & B1 & Wafer ii & Design G & 20 \\
    \cline{4-6}
      & & \multirow{2}{*}{B1} & \multirow{2}{*}{Wafer iii} & Design F & 38 \\
    \cline{5-6}
      & & & & Design G & 23 \\
    \hline
    \multirow{5}{*}{\textbf{Mask II}}
      & \multirow{2}{*}{n-type} & B2 & Wafer iv & Design H & 10 \\
    \cline{3-6}
      & & B4 & Wafer iv & Design I & 196 \\
    \cline{2-6}
      & \multirow{3}{*}{p-type} & B2 & Wafer vi & Design H & 61 \\
    \cline{3-6}
      & & \multirow{2}{*}{B3} & Wafer v & Design H & 19 \\
    \cline{4-6}
      & & & Wafer vii & Design H & 34 \\
    \hline
  \end{tabular}
\end{table}

\subsection*{CSDs Representative Examples}
\begin{figure}[h!]
    \centering
    \includegraphics[width=1\linewidth]{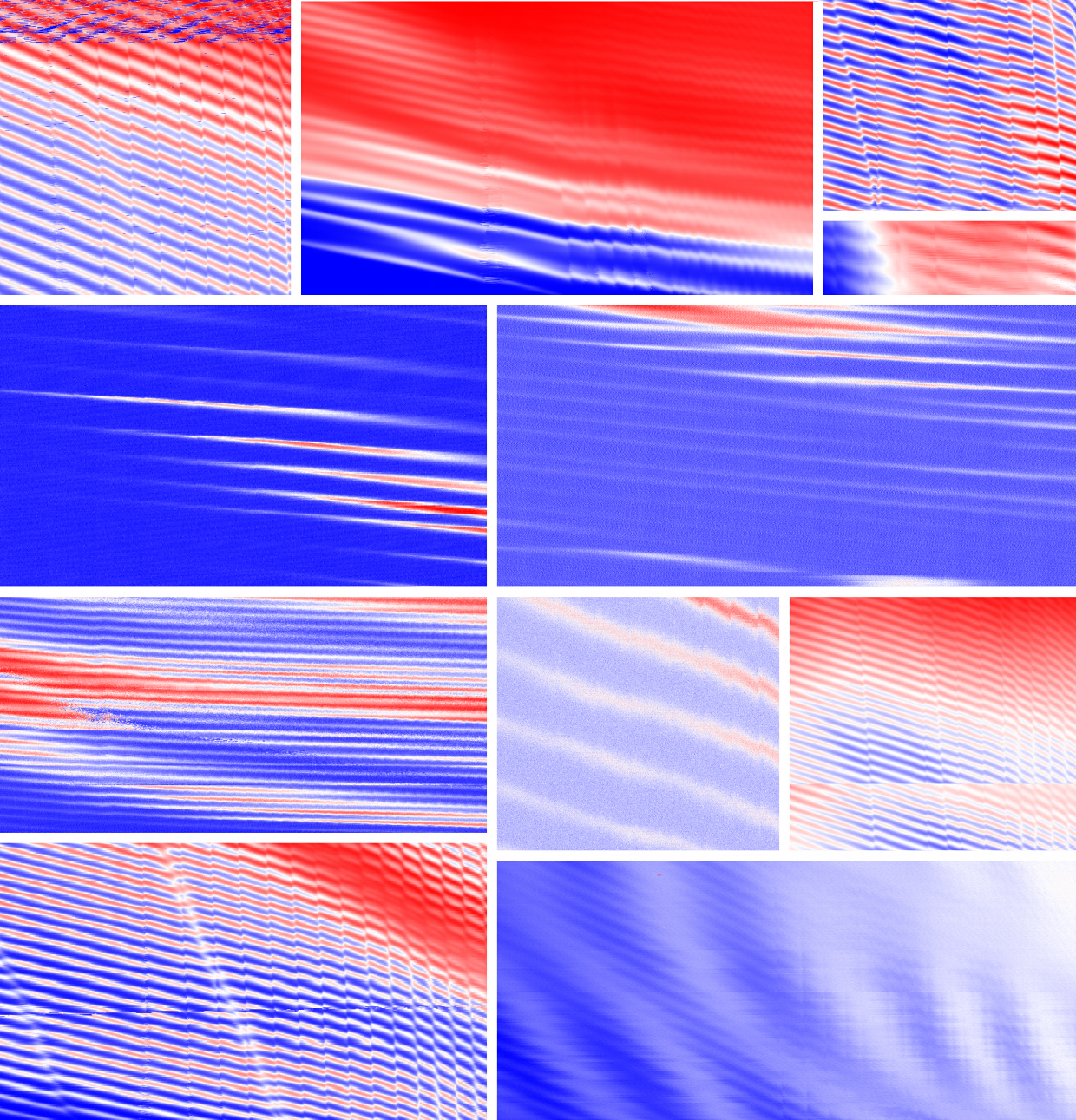}
    \caption{Representative stability diagram examples across masks, wafers and device geometries.}
    \label{fig:examples_CSD}
\end{figure}

To illustrate the variety and complexity of the collected data, we show some representative CSDs. Figure~\ref{fig:examples_CSD} displays examples spanning different masks, batches, and polarities. This diversity includes high signal-to-noise ratio (SNR) diagrams, those with faint transition lines, CSDs exhibiting stochastic noise, and those containing spurious lines, all of which reflect the realistic process variability and measurement challenges our model must handle.

\section{Labeling Process}

We developed an internal labeling tool, which consists of a GUI to annotate exported PNGs of each CSD and store annotations as polylines in a CSV. Labels were drawn with a fixed line width and adapted to each class of transitions differently. Every CSD was labeled by one annotator and checked by a second annotator. Ambiguous cases were flagged and resolved by consensus meetings. Final masks were rasterized from the CSV polylines to binary target masks \(Y(i,j)\in\{0,1\}\) and used as ground truth for training.

\section{Model Architecture and Training}
This section details the U-Net style segmentation model, its compact architecture, the specific hyperparameters used, and the training protocol implemented, including the rationale for grouped cross-validation to prevent any potential data leakage.

\subsection{Architecture}
We use a lightweight U-Net--style segmentation network that pairs a MobileNetV2 encoder (width multiplier $\alpha=1.4$) with a small custom decoder. Inputs are stability diagram images resized to $1024\times1024\times3$, and the network outputs a single per-pixel probability map
\[
\hat Y \in [0,1]^{1024\times1024}.
\]

\begin{figure}[h!]
    \centering
    \includegraphics[width=1\linewidth]{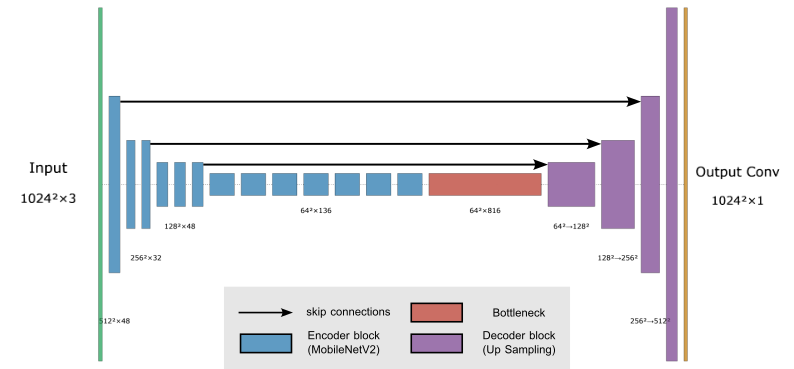}
    \caption{Schematic of the U-Net segmentation model (encoder, bottleneck, skip connections, and four-stage decoder) used to predict per-pixel transition probabilities from input CSD measurements.}
    \label{fig:architecture}
\end{figure}

An overview of the full network is shown in Figure~\ref{fig:architecture}.

\paragraph{Encoder (MobileNetV2)}
The encoder is a standard MobileNetV2 backbone initialized with ImageNet weights and fine-tuned during training. The choice of MobileNetV2 is due to its efficiency and compactness, which is highlighted in its architecture through:
\begin{itemize}
  \item \textbf{Depthwise separable convolutions} (spatial depthwise convs followed by $1\times1$ pointwise convs), which drastically reduce parameter count and multiply-accumulate operations compared to plain convolutions.
  \item \textbf{Inverted residual blocks with linear bottlenecks}, which preserve representational power while enabling narrow intermediate layers for efficiency.
\end{itemize}
We extract intermediate activations at four encoder resolutions to use as skip connections for the decoder.

\paragraph{Decoder}
The decoder performs a cascade of four upsampling stages that progressively restore spatial resolution. At each stage, the upsampled feature maps are concatenated with the corresponding encoder activation, followed by a small ConvBlock that refines features. Each ConvBlock implements:
\[
\text{ReflectionPad} \rightarrow \text{Conv2D}(3\times3) \rightarrow \text{BatchNorm} \rightarrow \text{ReLU}
\]
and is repeated twice per stage. Reflection padding reduces border artifacts that would otherwise create spurious short line fragments near image edges. Decoder channel widths are reduced progressively, and the final layer is a $1\times1$ convolution followed by a sigmoid activation producing $\hat Y$.

\paragraph{Loss function}
Because the segmentation task is highly class-imbalanced (thin transition lines vs. large background), we train with the Dice loss:
\[
  L_{\text{Dice}} = 1 - \frac{2\sum_i Y_i \hat Y_i + \epsilon}{\sum_i Y_i + \sum_i \hat Y_i + \epsilon},
\]
which directly optimizes overlap between prediction and ground truth ($Y$ is the binary ground truth, $\hat Y$ is the predicted probability map, and $\epsilon$ is a small constant for numerical stability). Training uses the Adam optimizer and the fixed hyperparameters listed in Table~\ref{tab:hyperparams_app}.

\paragraph{Model Size and Complexity}
Our model is compact, containing approximately 2.06 million parameters (about 7.85 MB as 32-bit floats), significantly fewer than many full U-Net variants. This modest parameter count reduces memory and storage requirements, making the model practical for repeated training and online deployment.

\begin{table}[!ht]
  \centering
  \caption{Model parameter count}
  \label{tab:model_summary}
  \begin{tabular}{lr}
    \toprule
    \textbf{Model} & \multicolumn{1}{c}{\textbf{Params}} \\
    \midrule
    MobileNetV2-U-Net (\(\alpha=1.4\)) (total params) & 2,057,489 \\
    Trainable params & 2,032,497 \\
    Non-trainable params & 24,992 \\
    \bottomrule
  \end{tabular}
\end{table}

\subsection{Training Protocol}
All experiments utilized the fixed hyperparameter set listed in Table~\ref{tab:hyperparams_app}. Training and large-scale inference were executed on $\text{NVIDIA V100}$ GPUs on the TGCC (Très Grand Centre de Calcul du $\text{CEA}$) cluster.

\begin{table}[h!]
\centering
\caption{Fixed hyperparameters used for all folds.}
\label{tab:hyperparams_app}
\begin{tabular}{|l|l|}
\hline
\textbf{Hyperparameter} & \textbf{Value} \\
\hline
Optimizer & Adam \cite{kingma2014adam} \\
\hline
Learning rate & $1\times10^{-4}$ \\
\hline
Loss function & Dice loss \\
\hline
Batch size & 8 \\
\hline
Epochs & 50 \\
\hline
Input size & 1024x1024 \\
\hline
MobileNetV2 width multiplier ($\alpha$) & $1.4$ \\
\hline
\end{tabular}
\end{table}

\paragraph{Group Cross-Validation}
We employed five-fold group cross-validation ($k=5$). The grouping key corresponds to the unique physical device identifier (which includes device design, die location, and gates used). This protocol guarantees that all CSDs from a single physical device are confined to a single fold (either training or test), thereby preventing data leakage between the training and test partitions. This ensures that a good test performance implies robustness to fabrication and measurement variability rather than memorization of device-specific patterns.

\section{Postprocessing and Auto-Tuning Algorithm}
This section describes the deterministic pipeline applied after the model produces a per-pixel probability map \(\hat Y\). The postprocessing sequence is designed to convert the predicted line maps into a gate-voltage recommendation to tune the device into the single-charge regime:

\begin{enumerate}[leftmargin=12pt,itemsep=3pt]
      \item \textbf{Forward pass:} Run the machine learning (ML) model on normalized, preprocessed input CSD \(X\) to obtain the pixel-wise prediction of transition lines \(\hat Y\in[0,1]^{H\times W}\).
  \item \textbf{Thresholding:} Binarize \(\hat Y\) using \(\tau = 0.75\). This selects high-confidence pixels as belonging to a transition line:
    \[
      \hat Y^\tau_{ij} = \begin{cases}1 & \hat Y_{ij}\ge\tau,\\ 0 & \text{otherwise.}\end{cases}
    \]
  \item \textbf{Morphological closing:} Apply a closing operation (dilation then erosion) with a vertical rectangular structuring element (empirical choice: \(\text{kernel} = 20\times 2\) px) to bridge small gaps and connect slightly fragmented vertical segments.
  \item \textbf{Connected-component filtering (dynamic):} Compute areas \(A_k\) of connected components. Compute mean area \(\bar A = \frac{1}{K}\sum_k A_k\). Remove components with area \(A_k < 0.75\times\bar A\). This dynamic threshold adapts to diagram scale and suppresses small spurious detections while keeping genuine transition lines.
  \item \textbf{Line ordering and selection:} Extract the remaining connected components as transition lines. For each component, compute its centroid and associated gate voltage (x coordinate value) and sort them by order. For n-type devices, select the two lowest-voltage (leftmost) detected lines; for p-type devices, select the two highest-voltage (rightmost) detected lines.
  \item \textbf{Single-charge polygon and center:} Form the polygonal region between the two selected lines. Compute the polygon center of mass \((i^*,j^*)\) and map to physical voltages \((V^*_{G1},V^*_{G2})\) using the known voltage axis ranges recorded during acquisition.
  \item \textbf{Success flag (offline validation):} Compare \((i^*,j^*)\) with the ground-truth single-charge mask (generated from the annotation of the transition lines) if the pixel falls within the ground-truth region, mark \texttt{Success} for offline evaluation.
\end{enumerate}

Then the original image coordinates \((x,y)\) are linearly mapped to gate voltages using the acquisition metadata (voltage per pixel), and the resulting voltage couple can be applied on the gates of the quantum dot (QD) for tuning.

\section{Results}
The model's performance is evaluated by its ability to successfully detect the single-charge regime after the full postprocessing pipeline is applied. Our evaluation uses the held-out test sets from the five-fold group cross-validation, ensuring that all test results are derived from devices unseen during training for that specific fold.

\begin{figure}[h!]
    \centering
    \includegraphics[width=1\linewidth]{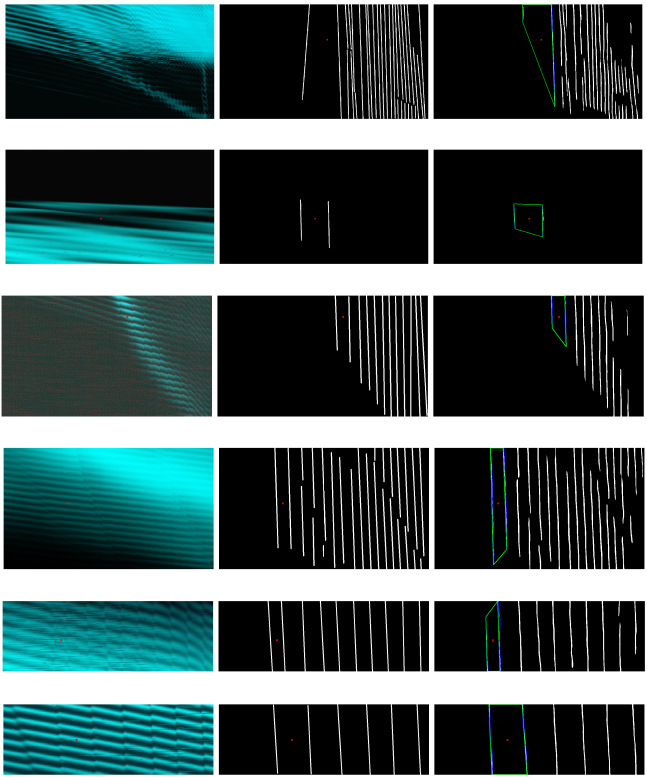}
    \caption{Examples of successful detection of the single charge regime across a variety of CSDs. Column 1 includes the original CSD used as input. Column 2 represents the corresponding hand-annotated ground truth mask. And column 3 is the model's prediction of transition lines and subsequent detection of the single charge regime by our auto-tuning pipeline.}
    \label{fig:supp:results}
\end{figure}

Figure~\ref{fig:supp:results} provides visual examples of the single-charge regime detection. The figure illustrates the full pipeline: starting from the raw CSD input (Column 1), comparing it to the human-annotated ground truth mask (Column 2), and concluding with the model's predicted line mask and the final single-charge region recommendation generated by our auto-tuning pipeline (Column 3). These examples confirm the model's ability to locate the target regime accurately across diverse diagrams.

Table~\ref{tab:per_fold_counts} summarizes the inference success rate, reporting the ratio of successful single-charge detections to the total number of diagrams for each design and each cross-validation fold. Overall, the model achieved a total success rate of $812/1015$ ($\mathbf{80.0\%}$) across all diverse devices and conditions. This demonstrates the robustness of the combined ML and deterministic postprocessing approach against real-world process and measurement variability.

\setlength{\tabcolsep}{1pt}
\begin{table}[h!]
  \centering
  \caption{\textbf{Inference Summary per Design and per Fold.} For each device and fold we report the ratio of the number of diagrams with successful single charge detection to the total number of diagrams, followed by the percentage value.}
  \label{tab:per_fold_counts}
  \begin{tabular}{lccccc}
    \toprule
    Fold & Design A & Design B & Design C & Design D & Design E \\
    \midrule
    1 & 11/12 (92\%)  & 14/16 (88\%)  & 17/17 (100\%) & 39/42 (93\%)  & 24/24 (100\%) \\
    2 & 17/19 (89\%)  & 10/11 (91\%)  & 15/20 (75\%)  & 21/26 (81\%)  & 37/41 (90\%)  \\
    3 & 11/15 (73\%)  & 7/12  (58\%)  & 12/15 (80\%)  & 25/30 (83\%)  & 25/31 (81\%)  \\
    4 & 12/14 (86\%)  & 9/9   (100\%) & 13/15 (87\%)  & 21/25 (84\%)  & 17/21 (81\%)  \\
    5 & 19/23 (83\%)  & 12/13 (92\%)  & 7/14  (50\%)  & 24/24 (100\%) & 19/21 (90\%)  \\
    \midrule
    Total & 70/83 (84\%) & 52/61 (85\%) & 64/81 (79\%) & 130/147 (88\%) & 122/138 (88\%) \\
    \bottomrule
  \end{tabular}

  \vspace{1ex}

  \begin{tabular}{lcccc}
    \toprule
    Fold & Design F & Design G & Design H & Design I \\
    \midrule
    1 & 17/18 (94\%) & 5/8  (62\%) & 19/29 (66\%) & 28/37 (76\%) \\
    2 & 23/24 (96\%) & 6/6  (100\%)& 8/20 (40\%)  & 25/36 (69\%) \\
    3 & 22/28 (79\%) & 5/9  (56\%) & 15/26 (58\%) & 32/37 (86\%) \\
    4 & 43/46 (93\%) & 6/9  (67\%) & 12/22 (55\%) & 28/42 (67\%) \\
    5 & 16/26 (62\%) & 8/11 (73\%) & 22/27 (81\%) & 34/44 (77\%) \\
    \midrule
    Total & 121/142 (85\%) & 30/43 (70\%) & 76/124 (61\%) & 147/196 (75\%) \\
    \bottomrule
  \end{tabular}

  \vspace{1ex}

  \noindent\textbf{Grand total (all shown devices)}: 812/1015 (80.0\%)
\end{table}

\section{Conclusion}
This supplementary document provides a comprehensive and technically detailed account of the methodology underlying our charge state auto-tuning algorithm. Specifically, we have showcased the variety of the 1015 CSD dataset, outlined the exact labeling protocol, detailed the architecture and compact nature of the MobileNetV2 U-Net architecture, and explained the use of a grouped cross-validation protocol to ensure robustness. Crucially, we have provided the complete, deterministic postprocessing algorithm, which acts as the final step to convert the predicted line masks into the actionable gate voltage values for the single-charge regime tuning. These details ensure the reproducibility and full understanding of the ML-driven quantum dot tuning approach presented in the main paper.

\bibliographystyle{apsrev4-2}
\bibliography{refs_apsrev4-2_2}
\end{document}